\documentclass[12pt]{article}

\usepackage[dvips]{graphicx}
\usepackage[english]{babel}
\usepackage{epsfig}
\usepackage{amsmath,amsfonts,amssymb,amsthm}
\usepackage{verbatim}
\usepackage{psfrag}
\usepackage{bm}
\usepackage{bbm}
\usepackage[square,comma,sort&compress,numbers]{natbib}
\usepackage{color}
\usepackage{slashed}
\usepackage{ulem}

\usepackage{epsf,epsfig}
\usepackage{graphics}
\usepackage{axodraw}
\usepackage{pstricks}

\begin{document}

\thispagestyle{empty}

\begin{flushright}
{
\small
TUM-HEP-875-13\\
TTK-13-02
}
\end{flushright}

\vspace{0.4cm}
\begin{center}
\Large\bf\boldmath
Right-Handed Neutrino Production at Finite Temperature:\\
Radiative Corrections, Soft and Collinear Divergences
\unboldmath
\end{center}

\vspace{0.4cm}

\begin{center}
Bj\"orn~Garbrecht$^\textnormal{\tiny\mbox{$\spadesuit$,$\diamondsuit$}}$, Frank Glowna$^\textnormal{\tiny\mbox{$\spadesuit$,$\diamondsuit$}}$ and Matti Herranen$^\textnormal{\tiny\mbox{$\clubsuit$,$\diamondsuit$}}$\\
\vskip0.2cm
{\it $^\textnormal{\tiny\mbox{$\spadesuit$}}$Physik Department T70, James-Franck-Stra{\ss}e,\\
Technische Universit\"at M\"unchen, 85748 Garching, Germany\\
\vspace{0.15cm}
$^\textnormal{\tiny\mbox{$\clubsuit$}}$Niels Bohr International Academy, Niels Bohr Institute and Discovery Center,\\
Blegdamsvej 17, DK-2100 Copenhagen, Denmark\\
\vspace{0.15cm}
$^\textnormal{\tiny\mbox{$\diamondsuit$}}$Institut f\"ur Theoretische Teilchenphysik und Kosmologie,\\
RWTH Aachen University, 52056 Aachen, Germany}\\
\vskip1.4cm
\end{center}

\begin{abstract}
The production and decay rate of massive sterile neutrinos at finite
temperature receives next-to-leading order corrections from the gauge interactions
of lepton and Higgs doublets. Using the Closed-Time-Path approach, we demonstrate
that the perturbatively obtained inclusive rate is finite. For this purpose, we
show that soft, collinear and Bose divergences cancel when adding the tree-level
rates from $1\leftrightarrow 3$ and $2\leftrightarrow 2$ processes
to vertex and wave-function corrections to $1\leftrightarrow 2$ processes.
These results hold for a general momentum of the sterile neutrino with respect
to the plasma frame. Moreover, they do not rely on non-relativistic approximations,
such that the full quantum-statistical effects are accounted for to the given order
in perturbation theory. While the neutrino production rate is of relevance for Leptogenesis,
the proposed methods may as well be suitable for application to a more
general class of relativistic transport phenomena.
\end{abstract}

\newpage

\tableofcontents

\newpage

\section{Introduction}

Scattering processes between particles at finite temperature
play an important role in
Early Universe Cosmology in view of applications such as Baryogenesis
and the production of Dark Matter. Besides, certain properties of hot and dense
strongly interacting matter, such as the photon production rate and
transport coefficients rely on the knowledge of these scattering rates.

It turns out useful to distinguish between the scatterings of four massless
particles, which are at leading order restricted to $2\leftrightarrow 2$
processes and the case where one of these particles is massive,
which opens up the phase space for $1\leftrightarrow 2$
and $1\leftrightarrow 3$ processes as well. In the case of the
scattering of massless
particles only, an important complication can arise from the $t$-channel exchange
of a massless particle, leading to a Coulomb divergence at tree level. This
divergence is mitigated by the screening in the plasma, and technically,
it can be removed by a resummation of thermal self-energies within the propagator
of the mediating particle~\cite{Arnold:2000dr,Arnold:2001ms,Besak:2012qm}.
When one of the scattering partners is massive,
this divergence no longer occurs, but instead, there are soft and collinear divergences.

The soft and collinear divergences are of the same type 
as {\it e.g.} those familiar from
QCD corrections to the quark pair production from an off-shell photon. While the
production rate for quark pairs and a soft or collinear gluon becomes non-perturbative
below a certain transversal momentum scale between the gluon and the emitting quark,
the inclusive cross section for pair production with and without gluons is
perturbatively well defined, due to the cancellation of soft and collinear divergences
between tree-level and loop diagrams. In the following, we refer to the soft and collinear
divergences collectively as infrared (IR) divergences.

It is natural to consider the same problem at finite temperature. Recently, this matter
has received attention in the context of
Leptogenesis~\cite{Besak:2012qm,Anisimov:2010gy,Salvio:2011sf,Laine:2011pq}.
The quantity of interest is the
relaxation rate of sterile
right-handed neutrinos $N$ toward equilibrium from decays and
inverse decays of Higgs bosons and leptons. When the mass $M$ of $N$ is small
compared to the temperature $T$, radiative corrections are of leading importance, because
the $1\leftrightarrow 2$ processes are kinematically suppressed. In the opposite case,
$M\gg T$, radiative corrections are subdominant to the leading
$1\leftrightarrow 2$ rates,
but yet, it may be interesting to accurately know the size of the next-to-leading order (NLO) corrections.
Furthermore, the situation $M\gg T$ corresponds to strong washout, which is in type-I see-saw scenarios perhaps favoured in the light of the observed mass-scale of the
active neutrinos.
Whether the $1\leftrightarrow 2$ processes are kinematically suppressed due to $M\ll T$
or not, we refer to the radiative corrections from Standard Model gauge interactions
as NLO in this paper.
The problem of the NLO corrections to $N$ production
for $M\gg T$ has recently been
resolved~\cite{Salvio:2011sf,Laine:2011pq}, making use of the fact that Maxwell statistics
is a good approximation in the non-relativistic
regime and that one may assume that the scale of the 
three momentum of $N$ is much below the temperature $T$.
Related to this matter are also works on the Drell-Yan production of dileptons
from a strongly interacting plasma~\cite{Baier:1988xv,Altherr:1989yn,Gabellini:1989yk,Altherr:1988bg},
and some useful calculational techniques are developed in Ref.~\cite{Laine:2011xm,Zhu:2012be},
where the spectral function of hot Yang-Mills theory is studied.

However, it still remains interesting to demonstrate the cancellation of
IR divergences between
tree-level and loop corrections at NLO, keeping full account
of relativistic effects and of quantum statistics. This is the main result
of this work. The proof of the cancellation of the IR divergences that
is presented here relies
partly on analytically isolating them from the integrals (for the wave-function
corrections) and partly on a rearrangement of the integrand such that only
integrable singularities remain (for the vertex corrections).
Therefore, this proof is already suggestive of
a method of numerically calculating the production rate, a task
that we will pursue in a separate work.

In order to compute the relaxation rate for $N$ at finite temperature, we
choose to use the Closed-Time-Path (CTP)
method~\cite{Schwinger:1960qe,Keldysh:1964ud}. In this formulation,
Schwinger-Dyson equations can be derived~\cite{Calzetta:1986cq,Prokopec:2003pj,Prokopec:2004ic}
(a particular subset of which are
known as the Kadanoff-Baym equations) that describe the time evolution
of the Quantum-Field-Theoretical non-equilibrium system. For this reason,
the CTP approach has recently been applied to Leptogenesis in different
parametric regimes, and various new effects and corrections have been derived~\cite{Buchmuller:2000nd,De Simone:2007rw,Garny:2009rv,Garny:2009qn,Anisimov:2010aq,Garny:2010nj,Beneke:2010wd,Beneke:2010dz,Garny:2010nz,Garbrecht:2010sz,Anisimov:2010dk,Garbrecht:2011aw,Garny:2011hg,Garbrecht:2012qv,Drewes:2012ma,Garbrecht:2012pq,Frossard:2012pc}
.
In the CTP formulation, the relaxation rate is directly related to the spectral
(or anti-hermitian) self energy.
As we assume that the fields that participate in gauge interactions are
in thermal equilibrium (which applies to all fields within the loop
diagrams that contribute to the self energy of $N$ in the present approximations),
we notice that the relaxation rate can be obtained from equilibrium
field theory as well.
Here, we choose the CTP approach because of its connection to non-equilibrium field theory
formulations for Leptogenesis. Moreover, the relation of the spectral self energy
to the relaxation
rate can be easily inferred from the kinetic equations that are derived from the
Schwinger-Dyson equations, ({\it cf.} Ref.~\cite{Beneke:2010wd}),
and finally, the CTP approach provides very efficient Feynman
rules that are perhaps easier to apply than certain
finite-temperature cutting rules used in Ref.~\cite{Salvio:2011sf}.

The methods that we employ in order to demonstrate the cancellation
of IR divergences are very explicit, which has the disadvantage
that they rely on a number of technical
details on the evaluation of phase-space
and loop integrals. On the other hand, the explicit demonstration
of the cancellation readily suggests a method of practical evaluation
of the neutrino relaxation rate to be pursued in a future work.
In order to communicate the main points of this work to a reader who
is not interested in the technical details and to give a first overview
to a reader who is at least partly interested in these details and their
reproduction, we provide in Section~\ref{section:summary} a
summary of our method and of the cancellation of the IR divergences,
that gives reference to the main results that are worked out in the technical
Sections. We begin Section~\ref{section:summary} with a part
containing general prerequisites and remarks
about the CTP approach and the
neutrino relaxation rate (Section~\ref{section:prerequisites}).
As suggested by the different diagrammatic contributions to the
relaxation rate, which also lead to different
calculational methods, we further present separate parts
about wave-function type (Section~\ref{Section:WV:overview})
and vertex-type (Section~\ref{section:VERT:overwiew}) corrections.
In Section~\ref{section:results},
we present the main results for the IR- and ultraviolet-finite neutrino
production rate, and the
the full calculational details are then presented in
Section~\ref{section:WV} for the wave-function corrections and
in Section~\ref{section:VERT} for the vertex corrections. We conclude
with Section~\ref{section:conclusions}.

\section{Summary of the Calculation and
Cancellation of Collinear, Soft and Bose Divergences}
\label{section:summary}

\subsection{Kinetic Equations, Spectral Self Energy and the Relaxation rate for Right-Handed Neutrinos}
\label{section:prerequisites}

We are interested in the production rate of right-handed neutrinos $N$,
or more precisely,
in their relaxation rate toward thermal
equilibrium. The self energy for $N$ is given by
$\slashed\Sigma_N$. At one-loop order, it originates from Yukawa interactions $Y$
(that we choose here to be real)
with lepton doublets $\ell$
and Higgs-boson doublets $\phi$.
Besides, $\slashed\Sigma_N$
receives higher order corrections due to the
${\rm SU}(2)$  and ${\rm U}(1)$ gauge interactions in the symmetric
Electroweak phase at high temperatures.
In particular, the model we consider is given by the Lagrangian
\begin{align}
{\cal L}={\cal L}_{\rm SM}+\frac 12 \bar\psi_N({\rm i}\slashed \partial -M)\psi_N
-Y \bar \psi_\ell \phi^\dagger P_{\rm R}\psi_N
-Y \bar \psi_N \phi P_{\rm L}\psi_\ell
\,,
\end{align}
where ${\cal L}_{\rm SM}$ is the Lagrangian of the Standard Model,
$\psi_{N,\ell}$ are the spinors associated with $N$ and $\ell$ and
$P_{\rm L,R}=(1\mp\gamma^5)/2$. (Group indices of ${\rm SU}(2)$ and
antisymmetric tensors that may be necessary to form invariant products are
suppressed throughout this paper.) On the spinor associated with
$N$, we impose the Majorana condition $\psi_N^c=C\bar\psi_N^T=\psi_N$,
where $C$ is the charge-conjugation matrix.
The gauge couplings are given by
by $g_2$ for ${\rm SU}(2)$ and $g_1$ for ${\rm U}(1)$, and moreover, we denote by
$g_w=2$ the dimension of the fundamental doublet representation of
${\rm SU}(2)$. As the Higgs doublet and the lepton doublet have the
same Electroweak charges, it is useful to define the recurring factor
\begin{align}
\label{G:Higgs:lepto}
G=\frac 34 g_2^2+\frac 14 g_1^2\,.
\end{align}
Besides, there are sizeable corrections from top-quark loops and
self-interactions of the Higgs boson. While we do not include
the latter explicitly, their treatment should
be straightforward, given the discussion of the wave-function
corrections in Sections~\ref{Section:WV:overview} and~\ref{section:WV}.

Since we assume the leptons, Higgs bosons
and gauge bosons to be in thermal equilibrium, the relaxation term
in the kinetic equation for the right handed neutrinos is proportional
to their deviation from equilibrium~\cite{Beneke:2010wd}:
\begin{align}
\label{kineq:N}
\frac{d}{dt}
f_{N}(\mathbf k)
=&\frac 14 \int \frac{d k_0}{2\pi}{\rm sign}(k_0) {\rm tr}\left[
{\rm i}{\Sigma\!\!\!/}_{N}^{>}(k) {\rm i}S_{N}^{<}(k)
-{\rm i}{\Sigma\!\!\!/}_{N}^{<}(k) {\rm i}S_{N}^{>}(k)
\right]
\\\notag
=&-\frac 12 \int \frac{d k_0}{2\pi}{\rm sign}(k_0) {\rm tr}\left[\slashed k
{\Sigma\!\!\!/}_{N}^{\cal A}(k)
\right]2\pi\delta(k^2-m_N^2)\delta f_N(\mathbf k)
\\\notag
=&-\frac{1}{2 k^0}
{\rm tr}\left[\slashed k\slashed\Sigma_{N}^{\cal A}(k)\right]
\delta f_N(\mathbf k)
\,,
\end{align}
where $k^0=\pm\sqrt{\mathbf{k}^2 +M^2}$ in the last term
and
where $\delta f_N(\mathbf k)$ is the deviation of the distribution function
$f_N(\mathbf k)$ of right-handed neutrinos from the Fermi-Dirac distribution.
We assume spatially homogeneous conditions and
work in Wigner space, where the two-point functions are Fourier transforms
with respect to the relative coordinate. Therefore, there remains a time
dependence, but we however
suppress the explicit time coordinate in the arguments of the Wigner 
functions and distribution functions.
As we calculate the self energies in this paper assuming thermal equilibrium
for the internal propagators, we make the
remark that the approach we use here is effectively identical to the real-time formulation of equilibrium field-theory. In Eq.~(\ref{kineq:N}),
we have made use of the definition of the spectral self-energy~(\ref{CTP:spectral})
and of the KMS relation for equilibrium two-point functions~(\ref{KMS}).
Notice that when $\slashed\Sigma_{N}^{<}(k)$ is known
in equilibrium, the KMS relations immediately yield
$\slashed\Sigma_{N}^{\cal A}(k)$ and therefore the relaxation rate.

In order to calculate $\slashed\Sigma_{N}^{\cal A}(k)$ when the mass $M$ of $N$
is non-zero, we perform a loop expansion. For this expansion to converge,
it is crucial that we calculate an inclusive production rate with or without the
production or absorption of a real gauge boson, such that IR-divergent contributions
from real emissions ($2\leftrightarrow2$ scatterings and $1\leftrightarrow 3$ decays and
inverse decays) and loop corrections (to the $1\leftrightarrow 2$ decays and
inverse decays) cancel. Note however that the perturbation
expansion breaks down in the limit $M\to 0$, because of the $t$-channel exchange
of massless fermions~\cite{Arnold:2000dr,Arnold:2001ms,Besak:2012qm}.
We will deal with this matter in a separate
publication~\cite{GSG}.

\begin{figure}[t!]
\begin{center}
\[
\textnormal{\scalebox{0.8}{  \begin{picture}(194,105) (215,-270)
    \SetWidth{1.0}
    \Line(216,-217)(264,-217)
\DashArrowArcn(312,-218.153)(48.014,1.376,-181.376){10}
    \Line(360,-217)(408,-217)
\ArrowArc(312,-216.571)(48.002,-0.512,180.512)
    \Vertex(264,-217){3}
    \Vertex(360,-217){3}
  \end{picture}}}
\;\;\raisebox{1.3cm}{\Large{+}}\;\;
\textnormal{\scalebox{0.8}{  \begin{picture}(194,105) (215,-270)
    \SetWidth{1.0}
    \Line(216,-217)(264,-217)
\DashArrowArc(312,-218.153)(48.014,-181.376,1.376){10}
    \Line(360,-217)(408,-217)
\ArrowArcn(312,-216.571)(48.002,180.512,-0.512)
    \Vertex(264,-217){3}
    \Vertex(360,-217){3}
  \end{picture}}}
\]
\end{center}
\caption{
\label{SigmaN:LO}
The self energy $[\slashed \Sigma_N]^{\rm LO}$.}
\end{figure}

The Yukawa coupling
between the right-handed neutrino $N$, the lepton doublet $\ell$ and
the Higgs doublet $\phi$ gives rise to the leading-order
contribution to the self energy of the right-handed neutrino, that
is represented diagrammatically in Figure~\ref{SigmaN:LO},
\begin{align}
\left[{\rm i}\slashed\Sigma_{N}^{ab}(p)\right]^{\rm LO}
=g_wY^2\int\frac{d^4 p}{(2\pi)^4}
\left(
{\rm i}S_\ell^{ab}(p+k){\rm i}\Delta_\phi^{ab}(-k)
+C[{\rm i}S_\ell^{ba}(-p-k)]^tC^\dagger{\rm i}\Delta_\phi^{ba}(k)
\right)
\,,
\end{align}
where $a,b=\pm$ are CTP indices (we follow the conventions used in
Ref.~\cite{Prokopec:2003pj})
and $t$ stands for a transposition of Dirac matrices.
(We indicate the various contributions to $\slashed\Sigma_{N}$ by superscripts
on square brackets, for the sake of readability. Notice that we deviate from
the conventions of Ref.~\cite{Prokopec:2003pj} in the detail that we
define the fermionic self-energies as $\slashed \Sigma$,
while in Ref.~\cite{Prokopec:2003pj} the same quantities are
defined without a slash.\footnote{Here it is understood that $\slashed \Sigma$ may involve
 chiral projection matrices $P_{\rm L,R}$, for instance for left-chiral fermions
 $\slashed \Sigma = P_{\rm R} \gamma^\mu\Sigma_\mu$.}) The tree-level propagators
for the lepton and the Higgs fields are given by Eqs.~(\ref{prop:ell:expl})
and~(\ref{prop:phi:expl}). Explicit expressions for
$\left[\slashed\Sigma_{N}^{\cal A}(p)\right]^{\rm LO}$ can be found
in Ref.~\cite{Beneke:2010wd}.

The main purpose of this work is to provide a method for calculating
the NLO contributions. These may be categorised
in two-particle-reducible wave-function type corrections (superscript WV, Figure~\ref{SigmaN:WV})
and two-particle-irreducible
vertex-type corrections (superscript VERT, Figure~\ref{SigmaN:VERT}).
Due to the $t$-channel divergences, it turns out that the two-particle-reducible contributions diverge for $M\to 0$, and instead, a two-particle-irreducible
one-loop diagram with resummed propagators should be calculated~\cite{GSG}.

In all our calculations, we denote by
\begin{align}
p^\mu=(M,0,0,0)
\end{align}
the momentum of the right-handed neutrino in the Centre of Mass System
(CMS).
The relative motion
with respect to the plasma is accounted for by a plasma vector
\begin{align}
\label{plasma:vector}
u^\mu=\frac1M(\tilde p^0,\tilde{\mathbf{p}})\,,
\end{align}
where $\tilde p^0$ and $\tilde{\mathbf{p}}$ are energy and three-momentum
of the right-handed neutrino in the plasma frame
(where the plasma is at rest).
Moreover we note that
\begin{align}
\label{Sigma:symm:neutral}
{\rm tr}[\slashed p\slashed \Sigma^{ab}_N(p)]={\rm tr}[\slashed p\slashed \Sigma^{ba}_N(- p)]\,,
\end{align}
which follows from the relation~(\ref{Majorana:selfenergy}),
such that our choice for $p$ with
$p^0>0$ does not imply a restriction of the results.

\subsection{Summary of Cancellation of IR Divergences in the Wave-Function Diagrams}
\label{Section:WV:overview}

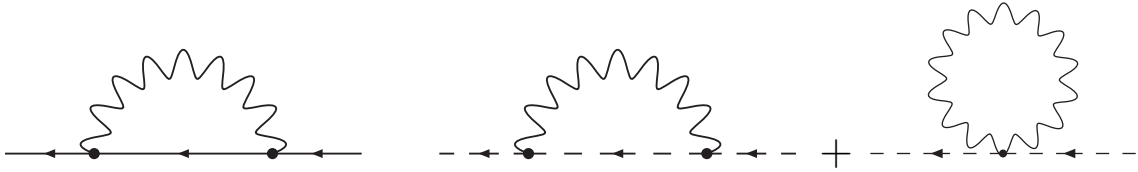
\begin{figure}[t!]
\begin{center}
\[
\textnormal{\scalebox{0.7}{  \begin{picture}(194,60) (215,-265)
    \SetWidth{1.0}
\ArrowLine(264,-262)(216,-262)
\ArrowLine(408,-262)(360,-262)
    \PhotonArc(312,-261.571)(48.002,-0.512,180.512){7.5}{8.5}
\ArrowLine(360,-262)(264,-262)
    \Vertex(360,-262){3}
    \Vertex(264,-262){3}
  \end{picture}}}
\;\;\phantom{\raisebox{-.1cm}{\Large{+}}}\;\;
\textnormal{\scalebox{0.7}{  \begin{picture}(194,60) (215,-265)
    \SetWidth{1.0}
\DashArrowLine(264,-262)(216,-262){10}
\DashArrowLine(408,-262)(360,-262){10}
    \PhotonArc(312,-261.571)(48.002,-0.512,180.512){7.5}{8.5}
\DashArrowLine(360,-262)(264,-262){10}
    \Vertex(360,-262){3}
    \Vertex(264,-262){3}
  \end{picture}}}
\;\;{\raisebox{-.1cm}{\Large{+}}}\;\;
\raisebox{.05cm}{\textnormal{\scalebox{1.0}{\begin{picture}(100,55)(0,0)
\DashArrowLine(100,0)(50,0){5}
\DashArrowLine(50,0)(0,0){5}
\Vertex(50,0){1.5}
\PhotonArc(50,28)(25,-97,263){3.75}{14}
\end{picture}}}}
\]
\end{center}
\caption{\label{SigmaellPiphi}
The amputated diagrams in this Figure represent  the gauge contributions to
$\slashed\Sigma_\ell$ (first diagram) and $\Pi_\phi$ (the sum of the second and third diagram).
When these diagrams are not amputated, they correspond to ${\rm i}S_\ell^{(1)}$ (first)
and ${\rm i}\Delta_\phi^{(1)}$ (sum of second and third). We refer to the first and second diagram as sunset diagrams, to the third as seagull diagram. Due to the CTP Feynman-rules,
the seagull diagram contributes to $\Pi_\phi^H$, but not to $\Pi_\phi^{\cal A}$.}
\end{figure}

The wave-function type contributions can be written as
\begin{align}
\left[\slashed\Sigma_{N}^{\cal A}(p)\right]^{\rm WV}
=g_w Y^2\int\frac{d^4 k}{(2\pi)^4}
\left[
{\rm i}S_\ell^{(1)>}(p+k){\rm i}\Delta_\phi^{>}(-k)
+{\rm i}S_\ell^{>}(p+k){\rm i}\Delta_\phi^{(1)>}(-k)
\right]-<\leftrightarrow>
\,,
\end{align}
where ${\rm i}S_\ell^{(1)}$ and ${\rm i}\Delta_\phi^{(1)}$ are the
one-loop corrections to the propagators, {\it i.e.} the self-energies
with external legs ({\it cf.} Figure~\ref{SigmaellPiphi}).
The particular diagrams contributing to $[\slashed\Sigma_{N}]^{\rm WV}$
are therefore two-particle
reducible, as apparent from the diagrammatic expression for
$[\slashed\Sigma_{N}]^{\rm WV}$ in
Figure~\ref{SigmaN:WV}.
It would be a simple matter to use fully resummed propagators rather
than the single-loop insertions, however it is interesting to verify that perturbation
theory is well defined without resorting to a resummation.

\begin{figure}[t!]
\begin{center}
\begin{align}
\notag
&\textnormal{\scalebox{0.8}{  \begin{picture}(194,132) (215,-243)
    \SetWidth{1.0}

    \Line(216,-190)(264,-190)
    \ArrowArc(308.5,-187.571)(44.566,-242.614,-176.876)
    \ArrowArc(312,-184.5)(43.684,56.674,123.326)
    \ArrowArc(315.438,-187.607)(44.627,-3.074,62.563)
    \DashArrowArcn(312,-191.153)(48.014,1.376,-181.376){10}
    \PhotonArc(312,-144.286)(24.286,-8.797,188.797){7.5}{4.5}
    \Line(360,-190)(408,-190)
    \Vertex(264,-190){3}
    \Vertex(360,-190){3}
    \Vertex(288,-149){3}
    \Vertex(336,-149){3}
  \end{picture}}}
\;\;\raisebox{1.3cm}{\Large{+}}\;\;
\textnormal{\scalebox{0.8}{  \begin{picture}(194,132) (215,-243)
    \SetWidth{1.0}

    \Line(216,-190)(264,-190)
    \ArrowArcn(308.5,-187.571)(44.566,-176.876,-242.614)
    \ArrowArcn(312,-184.5)(43.684,123.326,56.674)
    \ArrowArcn(315.438,-187.607)(44.627,62.563,-3.074)
    \DashArrowArc(312,-191.153)(48.014,-181.376,1.376){10}
    \PhotonArc(312,-144.286)(24.286,-8.797,188.797){7.5}{4.5}
    \Line(360,-190)(408,-190)
        \Vertex(264,-190){3}
    \Vertex(360,-190){3}
    \Vertex(288,-149){3}
    \Vertex(336,-149){3}
  \end{picture}}}
\\\notag
\;\;\raisebox{2.0cm}{\Large{+}}\;\;
&\textnormal{\scalebox{0.8}{  \begin{picture}(194,133) (215,-269)
    \SetWidth{1.0}
    \Line(216,-189)(264,-189)
    \Line(360,-189)(408,-189)
\DashArrowArcn(315.5,-191.429)(44.566,3.124,-62.614){10}
\DashArrowArcn(312,-194.5)(43.684,-56.674,-123.326){10}
\DashArrowArcn(308.5,-191.429)(44.566,-117.386,-183.124){10}
\ArrowArc(312,-187.847)(48.014,-1.376,181.376)
\PhotonArc(312,-235.857)(24.487,-191.441,11.441){7.5}{4.5}
    \Vertex(264,-190){3}
    \Vertex(360,-190){3}
    \Vertex(288,-231){3}
    \Vertex(336,-231){3}
  \end{picture}}}
\;\;\raisebox{2.0cm}{\Large{+}}\;\;
\textnormal{\scalebox{0.8}{  \begin{picture}(194,133) (215,-269)
    \SetWidth{1.0}

    \Line(216,-189)(264,-189)
    \Line(360,-189)(408,-189)
\DashArrowArc(315.5,-191.429)(44.566,-62.614,3.124){10}
\DashArrowArc(312,-194.5)(43.684,-123.326,-56.674){10}
\DashArrowArc(308.5,-191.429)(44.566,-183.124,-117.386){10}
\ArrowArcn(312,-187.847)(48.014,181.376,-1.376)
\PhotonArc(312,-235.857)(24.487,-191.441,11.441){7.5}{4.5}
    \Vertex(264,-190){3}
    \Vertex(360,-190){3}
    \Vertex(288,-231){3}
    \Vertex(336,-231){3}
  \end{picture}}}
\\\notag
\;\;\raisebox{3.0cm}{\Large{+}}\;\;
&\textnormal{\scalebox{1.54}{\begin{picture}(100,83)(0,0)
\Line(100,58)(75,58)
\Line(25,58)(0,58)
\ArrowArc(50,58)(25,0,180)
\DashArrowArcn(50,58)(25,0,90){5}
\DashArrowArcn(50,58)(25,90,180){5}
\Vertex(75,58){1.5}
\Vertex(25,58){1.5}
\PhotonArc(50,18)(12,-105,255){3.75}{6}
\Vertex(50,33){1.5}
\end{picture}}}
\;\;\raisebox{3.0cm}{\Large{+}}\;\;
\textnormal{\scalebox{1.54}{\begin{picture}(100,83)(0,0)
\Line(100,58)(75,58)
\Line(25,58)(0,58)
\ArrowArcn(50,58)(25,180,360)
\DashArrowArc(50,58)(25,270,360){5}
\DashArrowArc(50,58)(25,180,270){5}
\Vertex(75,58){1.5}
\Vertex(25,58){1.5}
\PhotonArc(50,18)(12,-105,255){3.75}{6}
\Vertex(50,33){1.5}
\end{picture}}}
\end{align}
\end{center}
\caption{\label{SigmaN:WV}
The self-energy term $[\slashed \Sigma_N]^{\rm WV}$. The first and second terms
constitute the contribution ${\cal F}$, the third to sixth the contribution
${\cal B}$ to ${\rm tr}[\slashed p\slashed \Sigma_N^{\cal A}]^{\rm WV}$.}
\end{figure}

The rate can be decomposed into a contribution ${\cal F}$
from gauge-boson radiation from the lepton and  a contribution ${\cal B}$
from gauge-boson radiation from the Higgs boson,
such that
\begin{subequations}
\begin{align}
&{\rm tr}\left[\slashed p \slashed\Sigma_{N}^{\cal A}(p)\right]^{\rm WV}={\cal F}+{\cal B}\,,
\\
\label{calF}
{\cal F}=&g_w Y^2\int\frac{d^4 k}{(2\pi)^4}
{\rm tr}\left[
\slashed p
\left(
{\rm i}S_\ell^{(1)>}(p+k){\rm i}\Delta_\phi^{>}(-k)
-{\rm i}S_\ell^{(1)<}(p+k){\rm i}\Delta_\phi^{<}(-k)
\right)
\right]
\,,
\\
{\cal B}=&g_w Y^2\int\frac{d^4 k}{(2\pi)^4}
{\rm tr}\left[
\slashed p
\left(
{\rm i}S_\ell^{>}(p+k){\rm i}\Delta_\phi^{(1)>}(-k)
-{\rm i}S_\ell^{<}(p+k){\rm i}\Delta_\phi^{(1)<}(-k)
\right)
\right]
\,.
\end{align}
\end{subequations}
The diagrams that represent $[\slashed \Sigma_N]^{\rm WV}$ and its decomposition in
${\cal F}$ and ${\cal B}$ are shown in Figure~\ref{SigmaN:WV}.

In order to explain how the cancellation of IR divergences works, we focus on emissions
of gauge radiation from the  Higgs boson, that is captured by
${\cal B}$. Radiation from the lepton can be treated
along the same lines, but is technically slightly more involved due to the
spinor structure. In Section~\ref{section:VERT}, we present all the necessary details on gauge radiation from the lepton. The one-loop correction to the scalar propagator
can be written as
\begin{subequations}
\begin{align}
{\rm i}\Delta_\phi^{(1)<,>}(k)=&2\Delta_\phi^{{\cal A}(1)}(k)
f_\phi^{<,>}(k\cdot u)\,,
\\
f_\phi^<(k\cdot u)=&f_\phi(k\cdot u)\,,\quad f_\phi^>(k\cdot u)=1+f_\phi(k\cdot u)\,,
\end{align}
\end{subequations}
where $\Delta_\phi^{{\cal A}(1)}(k)$ is the one-loop correction to the
spectral function, {\it cf.} Figure~\ref{SigmaN:LO} and Eqs.~(\ref{CTP:combinations}).
As we assume that thermal equilibrium is forced by the gauge interactions, throughout this
paper, we take for the distributions of Higgs bosons and gauge bosons $f_\phi$
and $f_A$ the Bose-Einstein distribution and for the distribution of leptons
$f_\ell$ the Fermi-Dirac distribution. Nonetheless, we keep the subscripts
$A$, $\phi$ and $\ell$ in order to indicate the origin of the individual distributions.

Now in thermal equilibrium, the one-loop correction to the spectral
function can be expressed as~\cite{Altherr:1994jc,Garbrecht:2011xw}
\begin{align}
\label{spectral:insert}
\Delta_\phi^{{\cal A}(1)}(k)=&
\frac12\left(
[\Delta_\phi^{R}(k)]^2[{\rm i}\Pi_\phi^{H}+\Pi_\phi^{\cal A}]
-[\Delta_\phi^{A}(k)]^2[{\rm i}\Pi_\phi^{H}-\Pi_\phi^{\cal A}]
\right)
\\\notag
=&\frac12\left([\Delta_\phi^{R}(k)]^2-[\Delta_\phi^{A}(k)]^2\right){\rm i}\Pi_\phi^{H}
+\frac12\left([\Delta_\phi^{R}(k)]^2+[\Delta_\phi^{A}(k)]^2\right)
\Pi_\phi^{\cal A}
\,,
\end{align}
where $\Pi_\phi^{{H},{\cal A}}$ are the hermitian and spectral self-energies of
the Higgs boson with a gauge boson in the loop. [{\it Cf.} Eqs.~(\ref{CTP:combinations})
for the definition of spectral and hermitian two-point functions.]
In principle, it
is important to assume here thermal equilibrium, because otherwise, there would
also occur the product of a retarded and an advanced propagator,
{\it i.e.} a pinch singularity, which leads to an ill-defined
integral because of a double pole
above and below the real axis. However, it is demonstrated in
Ref.~\cite{Altherr:1994jc} that a resummation of all loop insertions (in contrast
to the single loop insertion performed here) removes
the pinch singularities and leaves behind only the
equilibrium contributions to the propagator.
In Ref.~\cite{Garbrecht:2011xw}, it is shown that when performing the
gradient expansion, that occurs in the Wigner space formulation, to all orders,
the non-equilibrium part of the propagator takes the intuitively expected
resummed  form as well.
Recently, in Ref.~\cite{Millington:2012pf},
it has been proposed instead
to deal with this issue by using a two-momentum representation of the two-point function in contrast to the single-momentum
representation in Wigner space.

The retarded and advanced propagators that appear in Eq.~(\ref{spectral:insert})
are given by
\begin{align}
\label{retav}
{\rm i}\Delta_\phi^{R,A}=\frac{\rm i}{k^2\pm{\rm i}\,{\rm sign}k^0\varepsilon}
\,,
\end{align}
such that in the distributional sense,
\begin{subequations}
\label{retav:distr}
\begin{align}
\label{ret:distr}
[\Delta_\phi^{R}(k)]^2-[\Delta_\phi^{A}(k)]^2
=&-2\pi{\rm i}\sum\limits_{\pm}\delta(k^0\pm|\mathbf k|)
\left(
\frac{1}{4{k^0}^2}\frac{d}{d k^0}-\frac{1}{4{k^0}^3}
\right)\,,
\\
[\Delta_\phi^{R}(k)]^2+[\Delta_\phi^{A}(k)]^2
=&\frac{2}{k^4}\,.
\end{align}
\end{subequations}

Using above identities together with the expressions for the tree-level
propagators~(\ref{prop:phi:expl},\ref{prop:A:expl}),
the one-loop corrections due to real and virtual gauge bosons attaching
to the Higgs boson are
\begin{align}
\label{WB:boson:radiation}
{\cal B}=&g_w Y^2\int\frac{d^4k}{(2\pi)^4}
{\rm tr}
\left[
\slashed p{\rm i}S^>_\ell(p+k){\rm i}\Delta_\phi^{(1)>}(-k)
-\slashed p{\rm i}S^<_\ell(p+k){\rm i}\Delta_\phi^{(1)<}(-k)
\right]
\\
\notag
=&g_w Y^2\int\frac{d^3 k}{(2\pi)^3}\sum\limits_{k^0=-p^0\pm|\mathbf k|}
p^0 2\Delta_\phi^{(1){\cal A}}(k)
\left[f_\ell((p+k)\cdot u)+f_\phi(k \cdot u)\right]
\\\notag
=&g_w Y^2\int\frac{d^3 k}{(2\pi)^3}\sum\limits_{k^0=-p^0\pm|\mathbf k|}
p^0 \Bigg[2\pi\sum\limits_{\pm^\prime}\delta(k^0\pm^\prime|\mathbf k|)
\left(
\frac{1}{4{k^0}^2}\frac{d}{d k^0}-\frac{1}{4{k^0}^3}
\right)
\Pi_\phi^H(k)
\\\notag
+&
\frac{2}{k^4}\Pi_\phi^{\cal A}(k)\Bigg]
\times
\left[f_\ell((p+k)\cdot u)+f_\phi(k \cdot u)\right]
\,.
\end{align}
In order to identify the IR-divergent contributions, it is useful to
split the self-energy into parts that
are present for $T=0$ with a superscript ${\rm vac}$
and parts that vanish for $T=0$ with a superscript $T\not=0$:
\begin{align}
\Pi_\phi^{{H},{\cal A}}=\Pi_\phi^{{H},{\cal A},{\rm vac}}+\Pi_\phi^{{H},{\cal A},T\not=0}
\,,
\end{align}

The self-energies $\Pi_\phi^{H}$ and $\Pi_\phi^{\cal A}$ give rise to two different
contributions to ${\cal B}$. To those from $\Pi_\phi^{H}$, we refer to as
wave-function contributions, and to those from $\Pi_\phi^{\cal A}$ as
scattering contributions. Notice that the scattering contributions include besides
the tree-level $2\leftrightarrow2$ scattering rates the $1\leftrightarrow3$ decays
and inverse decays as well.

We now briefly
discuss the various contributions to ${\cal B}$ resulting
from this splitting and explain how the IR divergences present
in the particular contributions can be seen to cancel eventually:
\begin{itemize}
\item
\emph{Finite temperature contributions:}
The contributions from $\Pi_\phi^{{H},{\cal A},T\not=0}$ to
Eq.~(\ref{WB:boson:radiation}) lead to IR-finite integrals.
It is well known that the hermitian part of the thermal self-energy
 $\Pi_\phi^{{H}, T\not=0}$ is finite, {\it cf.} the explicit
expression~(\ref{PiH:Tnot=0}) below. Therefore, it gives rise
to a finite contribution
to the integral~(\ref{WB:boson:radiation}).
For the contribution from $\Pi_\phi^{{\cal A}, T\not=0}$, we
show that we can express
$\Pi_\phi^{{\cal A}, T\not=0}(k)=k^2 h(k)$, where $h(k)$ is a continuous
function, {\it cf.} Eqs.~(\ref{PiA:Tnot=0:expanded}) below. Therefore,
the first order singularity at $|\mathbf k|=k^0$ (or, equivalently,
$|\mathbf k|=p^0/2$) is integrable in the principal value sense.
\item
\emph{Vacuum wave-function contributions}:
It is well-known that the vacuum self-energy
$\Pi_\phi^{{H},{\rm vac}}$ for a scalar field with a gauge boson and a scalar boson
in the loop is IR-divergent. This divergence
can be regulated by a fictitious gauge-boson mass
$\lambda$. More precisely, the $\lambda$ dependence is $\propto\log\lambda$ and
given by Eqs.~(\ref{PiHvac},\ref{PiHvac:dimreg}) below.
\item
\emph{Real gauge boson emission (scatterings) in the vacuum}:
On the other hand, $\Pi_\phi^{{\cal A},{\rm vac}}(k)\propto k^2$, such that
the resulting singularity at $|\mathbf k|=k^0$ is again first order.
Now however, this gives an IR-divergent contribution to the integral,
because $\Pi_\phi^{{\cal A},{\rm vac}}(k)/k^2$ is
not continuous for $k^2=0$. Therefore, we introduce for this term the
gauge boson mass $\lambda$ as well.
The contribution from $\Pi_\phi^{{\cal A},{\rm vac}}$ to ${\cal B}$
adds to the rate of $1\to3$ particle decays of $N$.
Because $\Pi_\phi^{\cal A}(k)/k^0$ itself corresponds
to a decay rate of a scalar of mass-square $k^2$
into a massless scalar and a gauge boson of 
mass $\lambda$, there is a kinematic threshold that can be expressed in terms
of a Heaviside $\vartheta$-function, {\it cf.} Eq.~(\ref{PiA:vac}) below.
As a result, this contribution to the integral has an IR divergence
$\propto\log\lambda$ as well. When we perform an integration by parts, we
can isolate the IR divergence and compare it with the
one from wave-function corrections.
\item
\emph{Cancellation of IR divergences from wave-function and scattering corrections:}
Once the IR di\-vergences are isolated in terms $\propto\log\lambda$
by performing the integral over the $\delta$ function in Eq.~(\ref{WB:boson:radiation})
(for the virtual contributions from $\Pi_\phi^{{H},{\rm vac}}$)
or through integration by parts
(for the real contributions from $\Pi_\phi^{{\cal A},{\rm vac}}$), one
can see that the dependences on $\log\lambda$ cancel in the total result.
For this purpose, compare Eqs.~(\ref{B:virt:vac}) and~(\ref{B:real:vac+}).
This can be viewed as a consequence of the fact that
\begin{align}
\label{rel:spec:herm}
\vartheta(k^2-\lambda^2)
\pi\frac{d}{d\log\lambda^2}\Pi_\phi^{{H},{\rm vac}}(k)
={\rm sign}(k^0)\Pi_\phi^{{\cal A},{\rm vac}}(k)\,,
\end{align}
which is no accident, because $\Pi_\phi^{{H},{\rm vac}}$ and
$\Pi_\phi^{{\cal A},{\rm vac}}$ are real and imaginary part of the same
analytic self-energy.
\end{itemize}

The contributions from gauge-boson radiation from the lepton are
discussed in detail in Section~\ref{sec:rad:fermion}. For the
vacuum contributions to the spectral and hermitian self-energies,
the discussion goes along the same lines as for the radiation
from the Higgs boson. In addition, there are IR divergences
when the thermal part of the hermitian self-energy is evaluated on shell,
which are matched by IR-divergent contributions from
the thermal part of the spectral self-energy. We note here that
these thermal contributions which lead to IR divergences can be identified
with the self energies that one obtains in the hard thermal loop
(HTL) approximation. Otherwise,
the method of demonstrating
the cancellation of these divergences
is very similar to the one applied to the cancellation
of the IR-divergent vacuum-contributions.

\subsection{Summary of Cancellation of IR Divergences in the Vertex Diagram}
\label{section:VERT:overwiew}

The method relying on the relation~(\ref{rel:spec:herm}) between the spectral and hermitian self-energies
that we employ for the two-particle-reducible wave-function
contributions
obviously cannot be 
applied to the present case of the vertex diagram, Figure~\ref{SigmaN:VERT}.
One may again use a gauge-boson mass in order to regulate the
soft and collinear divergences. Due to the multiple integration boundaries that
appear for virtual corrections and real emissions,
one yet faces the problem of matching the particular contributions that
lead to a cancellation of the IR divergences, {\it i.e.} of the dependence
on the fictitious gauge-boson mass. This task can be facilitated by transforming the integrals
and by arranging and adding the particular integrands in such a manner that
there remains a single integrand that manifestly evaluates to a finite
answer. This is the method pursued in the present work.
One could then either go back to evaluate the particular divergent
contributions with a regulating gauge-boson mass, since it is then clear that
the dependence on the regulator eventually
cancels from the finite result. Alternatively,
one may directly integrate the total integrand without
introducing an IR regulator, what again leads to a finite 
result.

In order to accomplish the task of arranging the particular contributions
to the vertex-type self-energy into a manifestly IR-finite integral, we
focus on the reduced Wightman self-energy
${\rm tr}[\slashed p{\rm i}\slashed\Sigma_N^<(p)]$. This
quantity is proportional to the production rate of right-handed neutrinos
${\rm tr}[\slashed{\tilde p}{\rm i}\slashed\Sigma_N^<(\tilde p)]/(4\tilde{p}^0)$
(for $\tilde p^0=\sqrt{\mathbf{\tilde p}^2+M^2}$) [{\it cf.} Eq~(\ref{kineq:N}) and the definitions~(\ref{CTP:combinations})].
The thermal decay rate
${\rm tr}[\slashed{\tilde p}{\rm i}\slashed\Sigma_N^>(\tilde p)]/(4\tilde{p}^0)$
may be directly inferred
from the KMS relation~(\ref{KMS}),
${\rm i}\slashed\Sigma_N^>(p)=-{\rm e}^{\beta p\cdot u}{\rm i}\slashed\Sigma_N^<(p)$, and the
relaxation rate ${\rm tr}[\slashed{\tilde p}\slashed\Sigma_N^{\cal A}(\tilde p)]/(2{\tilde p}^0)$ can then
be found using the definitions~(\ref{CTP:combinations}).
Besides, we assume first that $p^0=M>0$, while the solution in the
case $p^0<0$ can be obtained from relation~(\ref{Sigma:symm:neutral}).

\begin{figure}[t!]
\begin{center}
\[
\textnormal{\scalebox{0.8}{  \begin{picture}(194,98) (239,-207)
    \SetWidth{1.0}
    \Line(240,-158)(288,-158)
    \Line(384,-158)(432,-158)
\ArrowLine(384,-158)(336,-110)
\ArrowLine(336,-110)(288,-158)
\DashArrowLine(384,-158)(336,-206){10}
\DashArrowLine(336,-206)(288,-158){10}
    \Photon(336,-110)(336,-206){7.5}{5}
        \Vertex(288,-158){3}
    \Vertex(384,-158){3}
    \Vertex(336,-110){3}
    \Vertex(336,-206){3}
  \end{picture}}}
\;\;\raisebox{1.2cm}{\Large{+}}\;\;
\textnormal{\scalebox{0.8}{  \begin{picture}(194,98) (239,-207)
    \SetWidth{1.0}
    \Line(240,-158)(288,-158)
    \Line(384,-158)(432,-158)
\ArrowLine(336,-110)(384,-158)
\ArrowLine(288,-158)(336,-110)
\DashArrowLine(336,-206)(384,-158){10}
\DashArrowLine(288,-158)(336,-206){10}
    \Photon(336,-110)(336,-206){7.5}{5}
    \Vertex(288,-158){3}
    \Vertex(384,-158){3}
    \Vertex(336,-110){3}
    \Vertex(336,-206){3}
  \end{picture}}}
\]
\end{center}
\caption{\label{SigmaN:VERT}
The self energy term $[\slashed \Sigma_N]^{\rm VERT}$.}
\end{figure}

The CTP expression for the 2-particle-irreducible vertex-type correction to the right-handed neutrino
self-energy, represented graphically in Figure~\ref{SigmaN:VERT}, is
\begin{align}
\left[{\rm i}\slashed\Sigma_N^{ab}\right]^{\rm VERT}(p)
=&2 g_w G Y^2\int\frac{d^4k}{(2\pi)^4}
\int\frac{d^4 q}{(2\pi)^4}
\sum\limits_{c,d}cd\,{\rm i}S_\ell^{ac}(k+q)\gamma^\mu{\rm i}\Delta^{cd}_{\mu\nu}(q)
\\\notag
\times&
\left[2(k-p)+q\right]^\nu
{\rm i}S_\ell^{cb}(k)
{\rm i}\Delta_\phi^{db}(-k+p)
{\rm i}\Delta_\phi(-k+p-q)
\,,
\end{align}
where the tree-level propagators are given by Eqs.~(\ref{prop:ell:expl},\ref{prop:phi:expl},\ref{prop:A:expl}).

\begin{figure}[t!]
\begin{center}
\[
\textnormal{\scalebox{0.8}{  \begin{picture}(194,114) (239,-199)
    \SetWidth{1.0}
    \Line(240,-142)(288,-142)
    \Line(384,-142)(432,-142)
    \Line(384,-142)(336,-94)
    \Line(336,-94)(288,-142)
\DashLine(384,-142)(336,-190){10}
\DashLine(336,-190)(288,-142){10}
    \Photon(336,-94)(336,-190){7.5}{5}
    \Line(352,-86)(352,-198)
        \Vertex(288,-142){3}
    \Vertex(384,-142){3}
    \Vertex(336,-94){3}
    \Vertex(336,-190){3}
  \end{picture}}}
\;\;\phantom{\raisebox{1.3cm}{\Large{+}}}\;\;
\textnormal{\scalebox{0.8}{  \begin{picture}(194,114) (239,-199)
    \SetWidth{1.0}
    \Line(240,-142)(288,-142)
    \Line(384,-142)(432,-142)
    \Line(384,-142)(336,-94)
    \Line(336,-94)(288,-142)
\DashLine(384,-142)(336,-190){10}
\DashLine(336,-190)(288,-142){10}
    \Photon(336,-94)(336,-190){7.5}{5}
    \Line(400,-86)(272,-198)
        \Vertex(288,-142){3}
    \Vertex(384,-142){3}
    \Vertex(336,-94){3}
    \Vertex(336,-190){3}
  \end{picture}}}
\]
\end{center}
\caption{\label{SigmaVERT:cut}Vertex ($[\slashed \Sigma_N]^{\rm vert}$)
and scattering ($[\slashed \Sigma_N]^{\rm sca}$)
contributions to $[\slashed \Sigma_N]^{\rm VERT}$.}
\end{figure}

The reduced
correction to the Wightman self-energy 
${\rm tr}[\slashed p  {\rm i}\slashed \Sigma_N^<(p)]^{\rm VERT}$
follows from
the $a,b=+,-$ contribution. Furthermore it is useful to shift in
the two terms where $c=-$ the momenta as $k\to k+q$ and $q\to-q$, to make
use of the cyclicity of the trace and of the definitions~(\ref{CTP:hermitian},\ref{CTP:statistic}), such that
one obtains
\begin{align}
\label{Sigmaless:start}
{\rm tr}[\slashed p  {\rm i}\slashed \Sigma_N^<&(p)]^{\rm VERT}
=4 g_w G Y^2\int\frac{d^4 k}{(2\pi)^4}\frac{d^4 q}{(2\pi)^4}
[2(k-p)+q]^\nu{\rm tr}\Big[\slashed p
\\\notag
\times\Big\{&{\rm i}S_\ell^H(k)\gamma^\mu
{\rm i}S_\ell^{<}(k+q){\rm i}\Delta_\phi^{<}(-k+p)
{\rm i}\Delta_{\mu\nu}^>(q){\rm i}\Delta^H_\phi(-k+p-q)
\\\notag
+&{\rm i}S_\ell^H(k)\gamma^\mu
{\rm i}S_\ell^<(k+q)
{\rm i}\Delta_\phi^<(-k+p-q)
{\rm i}\Delta_\phi^H(-k+p)
{\rm i}\Delta^F_{\mu\nu}(q)
\\\notag
+&{\rm i}S_\ell^H(k)\gamma^\mu
{\rm i}S_\ell^<(k+q)
{\rm i}\Delta_\phi^<(-k+p-q)
{\rm i}\Delta_\phi^F(-k+p)
{\rm i}\Delta^H_{\mu\nu}(q)
\\\notag
+&{\rm i}S_\ell^F(k)\gamma^\mu
{\rm i}S_\ell^<(k+q)
{\rm i}\Delta_\phi^<(-k+p-q)
{\rm i}\Delta_\phi^H(-k+p)
{\rm i}\Delta^H_{\mu\nu}(q)
\Big\}\Big]
\,.
\end{align}
Notice that the Wightman and statistic propagators (superscripts $<,>,F$) are
purely on shell, as they are proportional to an on-shell $\delta$ function, whereas
the hermitian propagators (superscript $H$) correspond to principal values.
It is therefore easy to see that
in this decomposition, the four terms in the integrand correspond
to the following rates:
\begin{itemize}
\item
The first term describes the interference of two tree-level
$2\leftrightarrow 2$ scattering or $1\leftrightarrow 3$ decay
diagrams, where in one of these, the gauge boson radiates from the lepton, in
the other one from the Higgs boson. We denote these contributions collectively
as scatterings (even though for $1\leftrightarrow 3$ processes this term does
not strictly apply) with the superscript sca,
${\rm tr}[\slashed p  {\rm i}\slashed \Sigma_N^<(p)]^{\rm sca}$. Diagrammatically,
this contribution corresponds to a cut through the Wightman propagators [those
with superscripts $<,>$, {\it cf.} Eq.~(\ref{CTP:prop:basic})], which
are purely on shell, {\it cf.} Figure~\ref{SigmaVERT:cut}.
\item
The second term yields a correction to the $1\leftrightarrow 2$ process.
It is the interference of the part of the vertex correction where
the gauge boson is on shell with the tree-level amplitude. We denote this contribution by
${\rm tr}[\slashed p  {\rm i}\slashed \Sigma_N^<(p)]^{{\rm vert}}_{1}$.
\item
The third term yields a correction to the $1\leftrightarrow 2$ process as well.
It is the interference of the part of the vertex correction where
the Higgs boson is on shell with the tree-level amplitude. We denote this contribution by
${\rm tr}[\slashed p  {\rm i}\slashed \Sigma_N^<(p)]^{{\rm vert}}_{2}$.
\item
The fourth term yields a correction to the $1\leftrightarrow 2$ process as well.
It is the interference of the part of the vertex correction where
the lepton is on shell with the tree-level amplitude. We denote this contribution by
${\rm tr}[\slashed p  {\rm i}\slashed \Sigma_N^<(p)]^{{\rm vert}}_{3}$.
\end{itemize}
The vertex corrections can be represented by cuts through the Wightman propagators as
well, {\it cf.} Figure~\ref{SigmaVERT:cut}.
We furthermore
emphasise that all of these individual contributions contain the full quantum statistical
factors for the on-shell particles.

\begin{figure}[t!]
\begin{center}
\epsfig{file=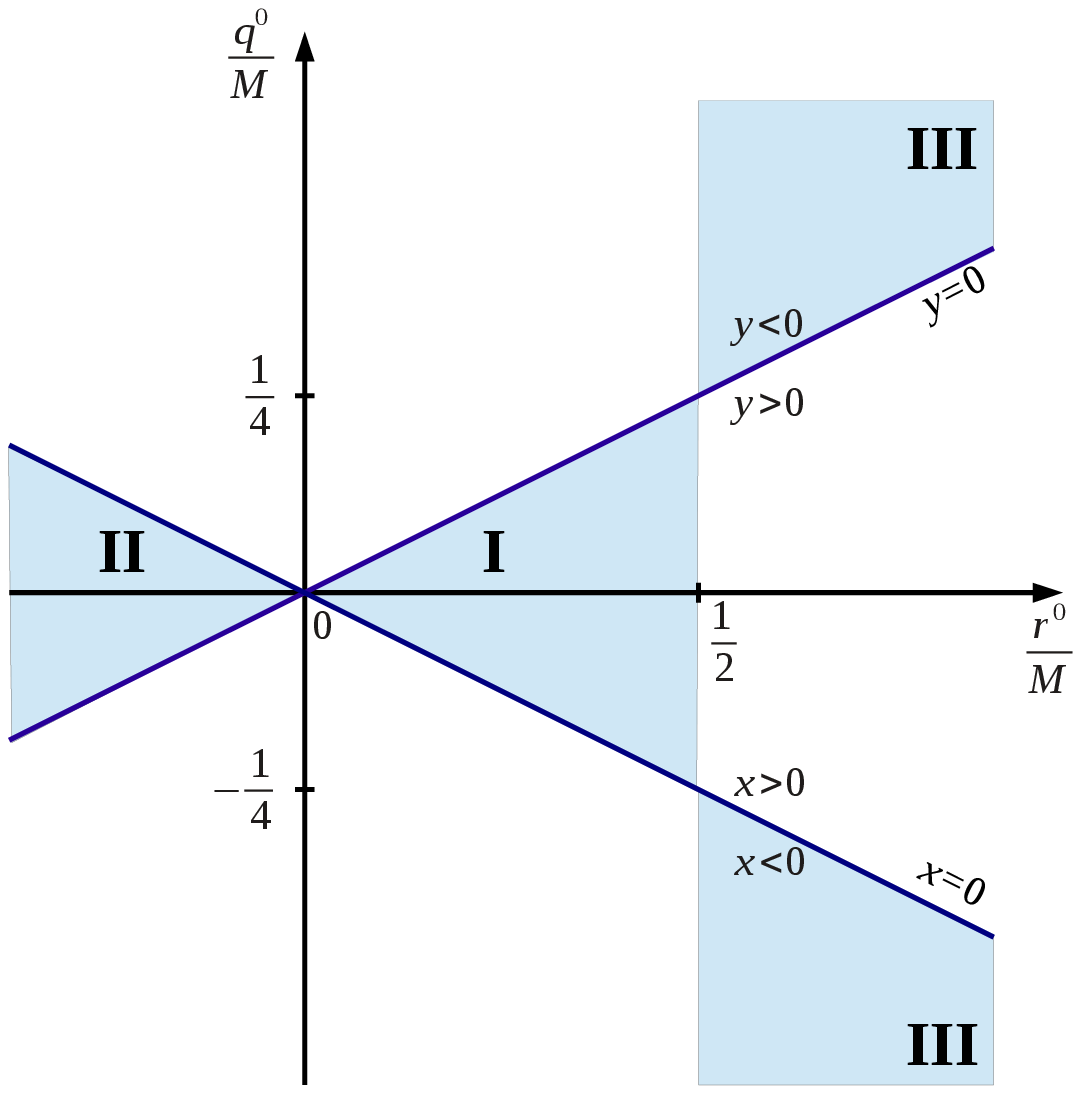,width=14.cm}
\end{center}
\caption{
\label{fig:kinregions}
Regions of integration for
${\rm tr}[\slashed p  {\rm i}\slashed \Sigma_N^<(p)]^{\rm sca}$, corresponding to tree-level $1\leftrightarrow 3$ and $2\leftrightarrow 2$
processes. The collinear singularities
coincide with the solid blue boundaries, $x=0$ or $y=0$.
Region~I corresponds to
$1\leftrightarrow 3$ processes, Region~II to the interference of two
$t$ channel amplitudes, the lower branch of Region~III to the interference
of one amplitude with a lepton in the $s$ channel with one with a Higgs
boson in the $t$ channel
and the upper branch of Region~III to the interference
of one amplitude with a lepton in the $t$ channel and one with a Higgs
boson in the $s$ channel.
}
\end{figure}

For all of these four individual contributions, IR divergences can occur
whenever the hermitian propagators go on shell. In order to identify the location
of the IR-divergent contributions to ${\rm tr}[\slashed p  {\rm i}\slashed \Sigma_N^<(p)]^{\rm sca}$, it is therefore useful
to parametrise the integrals by variables that are proportional to
the invariant momentum squares of these propagators. Here, we choose
$x=(-k+p-q)^2/M$ and $y=k^2/M$ with the momenta as parametrised
in Eq.~(\ref{Sigmaless:start}). [{\it Cf.} Eqs.~(\ref{xy:real}) for
the parametrisation in terms of the momenta chosen in Figure~\ref{fig:kinregions}.]
In addition, as we are working in a homogeneous
finite-temperature background, there eventually
remain two non-trivial angular integrations.
The kinematic
constraints that are forced within the CTP approach by the on-shell delta functions
within the Wightman and statistic propagators then imply that the integrand for
${\rm tr}[\slashed p  {\rm i}\slashed \Sigma_N^<(p)]^{\rm sca}$ has a limited
support, which is illustrated in Figure~\ref{fig:kinregions}. Notice that
the resulting sub-domains have simple kinematic interpretations that
are given in the Figure caption.
The lines where $x=0$ or $y=0$ correspond to the location of collinear
divergences, while the point where $x=y=0$ to the soft divergence,
where the emission of a zero-momentum gauge boson leads to a Sudakov logarithm square.
Because the collinear fringes where $x=0$ and $y=0$ are also the boundaries
of the integration domain for the real emissions, the singularities from
the hermitian propagators cannot be integrated in the principal value sense but lead
to logarithmic IR divergences. In addition, there are Bose singularities
from the gauge boson distribution at $x=y=0$ and from the Higgs boson distribution at
$(x,y)=(0,M/2)$.

These divergences cancel with IR divergences from the
corrections
${\rm tr}[\slashed p  {\rm i}\slashed \Sigma_N^<(p)]^{{\rm vert}}_{1,2,3}$. What we demonstrate in this work is that the integrals defining
${\rm tr}[\slashed p  {\rm i}\slashed \Sigma_N^<(p)]^{{\rm vert}}_{1,2,3}$
can be transformed
and added to ${\rm tr}[\slashed p  {\rm i}\slashed \Sigma_N^<(p)]^{\rm sca}$  in such a manner, that the resulting total integrand
in ${\rm tr}[\slashed p  {\rm i}\slashed \Sigma_N^<(p)]^{\rm VERT}$
has only integrable singularities after all. This procedure
is also suggestive of
a practical method for numerical integrations or analytical
approximations, that can be described as follows:
\begin{itemize}
\item
Parametrise the integral
${\rm tr}[\slashed p  {\rm i}\slashed \Sigma_N^<(p)]^{\rm sca}$
in terms of $x,y$ as described above and as illustrated in Figure~\ref{fig:kinregions} (Section~\ref{sec:vertex:real}).
\item
Parametrise the integrals
${\rm tr}[\slashed p  {\rm i}\slashed \Sigma_N^<(p)]^{{\rm vert}}_{1,2,3}$ by
$x$ and $y$ as well, such that these correspond to $1/M$ times the
momentum squares of the Higgs and the lepton propagators in the
sub-diagrams that correspond to virtual vertex corrections. Add various contributions
such that the integration area for the virtual corrections
covers the quadrant $x>0$, $y>0$ (Sections~\ref{sec:vertex:virt:1},~\ref{sec:vertex:virt:2} and ~\ref{sec:vertex:virt:3}).
\item
For the real corrections, fold Regions~II and~III (either or both $x,y<0$) to the quadrant where
$x,y>0$.
\item
Now add all contributions. The resulting integrand
${\cal J}^{\rm total}(x,y)$, Eq.~(\ref{integrand:IR:finite}),
only contains integrable
singularities (Section~\ref{sec:IR:finite}).
\item
The integral is not yet convergent for large values of $x$, $y$, what
corresponds to an ultraviolet (UV) divergence. We obtain a convergent
integral by subtracting a term that corresponds to the vertex correction
to the vacuum decays $N\to \ell\phi$ weighted by the thermal
distributions for $\ell$ and $\phi$, $\bar{\cal J}^{\rm vert,vac}$,
Eq.~(\ref{J:vert:vac}).
As this contribution is IR divergent, we cancel it by adding a correspondingly weighted rate for $1\to 3$
decays, $\bar{\cal J}^{\rm sca,vac}$,
Eq.~(\ref{J:sca:vac}). The subtracted contributions must be added again to the final 
results, but for these particular terms,
we can perform the integration over $dx\,dy$ analytically
and thus isolate and renormalise the UV divergence
(Section~\ref{sec:vert:UV}).
\end{itemize}

The point stating the presence of integrable singularities only
requires a proof, that we present
in Section~\ref{section:VERT}
and that at least in parts elucidates
how the cancellation of IR divergences
works in the finite-temperature background. The various
contributions to the integrand obviously factorise into kinematic and quantum statistical
terms. We perform an expansion of the kinematic factors and the arguments of the 
statistical functions that applies to the collinear fringes, where either
$|x|\ll M$ or $|y|\ll M$ as well as close to the point of the soft divergence,
where both $|x|,|y|\ll M$. Using these expansions, in Section~\ref{sec:IR:finite},
we demonstrate the following points:
\begin{itemize}
\item
On the collinear fringes, where $x=0$ or $y=0$, the total integrand takes finite values
and is hence integrable, as it is expressed by Eqs.~(\ref{cancel:x})
and~(\ref{cancel:y}).
\item
The point $x=y=0$ requires special care, since the soft gauge-boson singularity coincides
with the Bose divergence of the gauge-boson distribution function. We show that
the integrand behaves $1/\sqrt{x^2+y^2}$ for $(x,y)\to 0$, such that this isolated
singularity is integrable as well.
A technical point is here that contributions from different regions of the angular
integrations must be averaged.
\item
The divergence from the Higgs boson distribution functions is integrable in the principal 
value sense.
\end{itemize}
In order to show the cancellation of the singularities on the collinear fringes,
we need to assume that Higgs bosons, leptons and gauge bosons are in thermal equilibrium.
As a consequence, the collinear splitting processes are in equilibrium as
well, what leads to the detailed balance relations~(\ref{detailed:balance}), that are
essential in order to show that the integrand is finite on the fringes.

\section{NLO Neutrino Production Rate}
\label{section:results}

In this Section, we present the final results for the neutrino relaxation rate in terms of manifestly IR-
 and UV-finite integrals. The explicit expressions for the various terms are given in the subsequent Sections.
 The total relaxation rate in the plasma frame is [{\it cf.} Eq.~(\ref{kineq:N})]
\begin{align}
\frac{1}{2\tilde p^0}{\rm tr}\left[\slashed{\tilde p}\slashed\Sigma_{N}^{\cal A}(\tilde p)\right] = \frac{1}{2\tilde p^0}\left(
{\rm tr}\left[\slashed {\tilde p}\slashed\Sigma_{N}^{\cal A}(\tilde p)\right]^{\rm LO}
+{\rm tr}\left[\slashed {\tilde p}\slashed\Sigma_{N}^{\cal A}(\tilde p)\right]^{\rm WV}
+{\rm tr}\left[\slashed {\tilde p}\slashed\Sigma_{N}^{\cal A}(\tilde p)\right]^{\rm VERT}\right)\,.
\end{align}
Expressions for the leading order term
$\left[\slashed\Sigma_{N}^{\cal A}(\tilde p)\right]^{\rm LO}$ can be found
in Ref.~\cite{Beneke:2010wd}.
The wave-function-type contributions are given by\footnote{Due to Lorentz covariance: ${\rm tr}\left[\slashed {\tilde p}\slashed\Sigma_{N}^{\cal A}(\tilde p)\right] = {\rm tr}\left[\slashed p\slashed\Sigma_{N}^{\cal A}(p)\right]$.}
\begin{align}
\label{result:wv}
{\rm tr}\left[\slashed p\slashed\Sigma_{N}^{\cal A}(p)\right]^{\rm WV}=&\left({\cal B}^{{\rm wv},{\rm vac}}+{\cal B}^{{\rm sca},{\rm vac}}\right)
+{\cal B}^{{\rm wv},T\not=0}+{\cal B}^{{\rm sca},T\not=0}
\\\notag+&
\left({\cal F}^{{\rm vac},{\rm col},{\rm wv}}+{\cal F}^{{\rm vac},{\rm col},{\rm sca}}\right)
+\left({\cal F}^{{\rm HTL},{\rm col},{\rm wv}}+{\cal F}^{{\rm HTL},{\rm col},{\rm sca}}\right)
\\\notag+&
{\cal F}^{\rm vac,fin}+{\cal F}^{{\rm HTL},{\rm fin}}+{\cal F}^{T\not=0}\,,
\end{align}
and where the ${\cal B}$ terms correspond to radiation from the Higgs boson and ${\cal F}$ to radiation from the fermion.
 The various wave function, scattering, vacuum, collinear and HTL (sub)contributions
 are isolated and collected in a way that the IR divergences cancel within each
of the parentheses. The UV divergences
 cancel among the vacuum parts against the counter terms. The expressions for the various ${\cal B}$ and ${\cal F}$
 contributions are given in Section~\ref{section:WV}. 

The vertex-type contribution to the relaxation rate is given by
\begin{align}
\label{result:vertex:1}
f_{N}^{\rm eq}(p \cdot u){\rm tr}\big[\slashed p \slashed \Sigma_N^{\cal A}(p)\big]^{\rm VERT}
=&-\frac{2 g_w G Y^2}{32(2\pi)^4}\int\limits_0^{2\pi} d(\varphi-\psi)\int\limits_{-1}^{1}d\cos\varrho
\int\limits_0^\infty dx \int\limits_0^\infty dy{\cal J}^{\rm total}_{\overline{\rm UV}}(x,y)
\notag\\\notag
&-2 g_w G Y^2 \frac{1}{32(2\pi)^3}
\int\limits_{-1}^{1}d\cos\varrho
\left(
\bar{\cal J}^{\rm vert,vac}+\bar{\cal J}^{\rm sca,vac}
\right)
\\
&-\frac{M^2}{8\pi}Y\delta Y\int\limits_{-1}^{1}d\cos\varrho\,f_\ell(E_+)f_\phi(E_-)
\,,
\end{align} 
with
\begin{align}
\label{result:vertex:2}
{\cal J}^{\rm total}_{\overline{\rm UV}}(x,y)
=\frac14\Big(
&{\cal J}^{\rm sca}(x,y)+{\cal J}^{\rm sca}(-x,-y)+{\cal J}^{\rm sca}(-x,y)+{\cal J}^{\rm sca}(x,-y)
\\\notag+&
{\cal J}_{1+}^{\rm vert}(x,y)+{\cal J}_{1-}^{\rm vert}(x,y)+{\cal J}_{2+}^{\rm vert}(x,y)+{\cal J}_{2-}^{\rm vert}(x,y)
\\[2mm]\notag+&
{\cal J}_{3+}^{\rm vert}(x,y)+{\cal J}_{3-}^{\rm vert}(x,y)
\\\notag-&
{\cal J}^{\rm sca,vac}(x,y)-{\cal J}^{\rm vert,vac}(x,y)
+x\leftrightarrow y
\Big)+\varphi\to\varphi+\pi\,,
\end{align}
where the 
${\cal J}^{\rm sca}(x,y)$ terms correspond to $2\leftrightarrow 2$ scatterings and $1\leftrightarrow 3$ decay and inverse decay
 processes with the kinematic regions\footnote{We have recast the integration area to positive $x,y > 0$,
 hence the explicit minus signs in the arguments of contributions corresponding to regions II and III.} indicated in
 Figure~\ref{fig:kinregions}, and ${\cal J}_{1\pm}^{\rm vert}(x,y)$, ${\cal J}_{2\pm}^{\rm vert}(x,y)$ and
 ${\cal J}_{3\pm}^{\rm vert}(x,y)$ correspond to vertex corrections with an on-shell gauge boson, Higgs boson and lepton,
 respectively. The vacuum-type contributions ${\cal J}^{\rm sca,vac}(x,y)$ and ${\cal J}^{\rm vert,vac}(x,y)$ are subtracted
 within ${\cal J}^{\rm total}_{\overline{\rm UV}}(x,y)$ to make it UV finite (in addition to being manifestly IR finite), and
 they are cancelled by the vertex counter-term $\delta Y$. The Fermi-Dirac distribution $f_{N}^{\rm eq}$ in the front of Eq.~(\ref{result:vertex:1}) results
 from using the KMS relation~(\ref{KMS}). The explicit expressions for the integrands are presented in
 Section~\ref{section:VERT}. In addition to the kinematic $x,y$-variables, the integrands depend non-trivially also on
 the angles $\varphi-\psi$ and $\rho$ depicting the orientation between loop momenta and the plasma vector $\mathbf{u}$, and $E_\pm$ are given by Eq.~(\ref{E:pm}).

\section{Wave-Function Contributions}
\label{section:WV}

\subsection{Radiation from the Higgs Boson}

We now present the technical details that are omitted in
Section~\ref{Section:WV:overview}, where a more
qualitative overview of the
present approach to the regulation and cancellation of the IR divergences
is provided.

For a massless scalar radiating a gauge boson with the IR-regulating
mass $\lambda$, the vacuum and finite-temperature
contributions to
the hermitian self-energy are
\begin{subequations}
\begin{align}
\Pi_\phi^{H,{\rm vac}}(p)
=&-\frac{G}{2}\int\frac{d^4 q}{(2\pi)^4}\int\frac{d^4 k}{(2\pi)^4}
(2\pi)^4\delta^4(p-k-q)(p+k)^2
\\\notag
\times&
\left[
{\rm PV}\frac{1}{k^2}2\pi\delta(q^2-\lambda^2)
+{\rm PV}\frac{1}{q^2-\lambda^2}2\pi\delta(k^2)
\right]
\\\notag
+&
4 G\int\frac{d^4 k}{(2\pi)^4} \frac{\rm i}{k^2-\lambda^2+{\rm i}\varepsilon}
\,,
\\
\Pi_\phi^{{H},T\not=0}(p)
=&-G\int\frac{d^4 q}{(2\pi)^4}\int\frac{d^4 k}{(2\pi)^4}
(2\pi)^4\delta^4(p-k-q)(p+k)^2
\\\notag
\times&
\left[
{\rm PV}\frac{1}{k^2}2\pi\delta(q^2-\lambda^2)f_A(|q\cdot u|)
+{\rm PV}\frac{1}{q^2-\lambda^2}2\pi\delta(k^2)f_\phi(|k\cdot u|)
\right]
\\\notag
+&
4 G\int\frac{d^4 k}{(2\pi)^4}2\pi\delta(k^2-\lambda^2)f_A(k^0)
\,,
\end{align}
\end{subequations}
where $G$ as defined by Eq.~(\ref{G:Higgs:lepto})
encompasses the gauge coupling constants
and $u$ is the plasma vector defined in Eq.~(\ref{plasma:vector}). The first of the
integrals for $\Pi_\phi^{{H},{\rm vac}}$ and $\Pi_\phi^{{H},T\not=0}$ correspond to
the sunset diagrams, the second of the integrals to the
seagull diagrams ({\it cf.} Figure~\ref{SigmaellPiphi}).

A possible way of evaluating $\Pi_\phi^{H,{\rm vac}}$ is to go to
the frame where $\mathbf p=\mathbf 0$ and to introduce a momentum cutoff $\Lambda$.
Then,
\begin{subequations}
\begin{align}
\frac{\partial\Pi_\phi^{H,{\rm vac}}(p^0,\mathbf 0)}{\partial (p^0)^2}
&=\frac{G}{4\pi^2}\left(\log\frac{\Lambda}{\lambda}+\log2+\frac58\right)\,,
\\
\label{PiHvac}
\Rightarrow
\Pi_\phi^{H,{\rm vac}}(p)&=\frac{G}{4\pi^2}p^2\left(\log\frac{\Lambda}{\lambda}+C_1\right)+C_2+p^2\delta Z_\phi +\delta m_\phi^2\,,
\end{align}
\end{subequations}
where $\delta Z_\phi$ and $\delta m_\phi^2$ are field-strength
and mass counterterms.
Alternatively, of course, an effectively equivalent
result may be obtained using other regularisation
procedures, {\it e.g.} dimensional regularisation in $4-\epsilon$ dimensions,
such that the leading terms in $p^2$ are
\begin{align}
\label{PiHvac:dimreg}
\Pi_\phi^{H,{\rm vac}}(p)&=\frac{G}{4\pi^2}p^2\left(\log\frac{\mu}{\lambda}+\frac12 \Delta_\epsilon+\frac12\right)+p^2\delta Z_\phi\,,
\end{align}
where
\begin{align}
\Delta_\epsilon=\frac2\epsilon-\gamma_{\rm E}+\log(4\pi).
\end{align}
Notice that, as it is well known, the seagull graph vanishes when evaluated
using dimensional regularisation.
For both regularisation procedures, the UV divergences should be cancelled by the
field-strength renormalisation $\delta Z_\phi$. In the following,
we effectively account for this by replacing $\Lambda\to \bar\Lambda$,
which takes a finite, renormalisation-scheme dependent value.

For the part that vanishes as we take $T\to 0$, we set $\lambda=0$ and
obtain
\begin{align}
\label{PiH:Tnot=0}
\Pi_\phi^{{H},T\not=0}(p)=&G\frac{T^2}{8}-G\frac{p^2}{4\pi^2|\mathbf p|}
\\\notag
\times&\int d|\mathbf k|
\bigg[
\log\left|\frac{p^2-2|\mathbf k|p^0+2|\mathbf k||\mathbf p|}{p^2-2|\mathbf k|p^0-2|\mathbf k||\mathbf p|}\right|
+\log\left|\frac{p^2+2|\mathbf k|p^0+2|\mathbf k||\mathbf p|}{p^2+2|\mathbf k|p^0-2|\mathbf k||\mathbf p|}\right|
\bigg]f_{\rm B}(|\mathbf k|)
\,.
\end{align}
This is IR finite, such that it is indeed justified to take $\lambda=0$ for this
contribution.

For the spectral self-energy, we perform the split into
vacuum and finite-temperature parts
accordingly. Notice that due to the CTP Feynman-rules, the spectral self-energy
only receives sunset and no seagull contributions. The zero-temperature part
is
\begin{align}
\label{PiA:vac}
\Pi_\phi^{{\cal A},{\rm vac}}(k)=&-G\frac{2k^2-\lambda^2}{16\pi}\frac{k^2-\lambda^2}{k^2}\vartheta(k^2-\lambda^2){\rm sign}k^0
\approx
-G\frac{1}{8\pi}k^2\vartheta(k^2-\lambda^2){\rm sign}k^0\,,
\end{align}
where for the approximation, we neglect numerator terms of order $\lambda^2$.
The Heaviside $\vartheta$-function occurs due to the mass threshold of the would-be process
of a
scalar boson of mass square $k^2$ decaying into a massless scalar boson and a gauge boson of
mass $\lambda$. (The minus sign in front of the rate is not problematic, because
an on-shell scalar particle cannot actually
change its mass by radiating a gauge boson when the gauge symmetry
is unbroken.)
For the finite temperature contribution, we obtain
\begin{subequations}
\label{Pi:spectral:phi}
\begin{align}
\Pi^{{\cal A},T\not=0}_\phi(k)
=&\frac{G}{8\pi}\frac{k^2}{|\mathbf k|}
\left(
2 |\mathbf k|
-\frac{2}{\beta}\log
\frac{1-{\rm e}^{\beta\frac{k^0+|\mathbf k|}{2}}}
{1-{\rm e}^{\beta\frac{k^0-|\mathbf k|}{2}}}
\right)
\;\;\textnormal{for}\;\;k^2\geq0\,,
\\
\Pi^{{\cal A},T\not=0}_\phi(k)
=&\frac{G}{8\pi}\frac{k^2}{|\mathbf k|}
\left(
2k^0
-\frac{2}{\beta}\log
\frac{1-{\rm e}^{\beta\frac{|\mathbf k|+k^0}{2}}}
{1-{\rm e}^{\beta\frac{|\mathbf k|-k^0}{2}}}
\right)
\;\;
\textnormal{for}\;\;k^2<0\,.
\end{align}
\end{subequations}

We aim to calculate the correction ${\cal B}$ [defined in Eq.~(\ref{WB:boson:radiation})] by gauge-boson emission
from the Higgs boson to the relaxation rate of right-handed neutrinos.
The term that depends on $\Pi_\phi^{H}$ can
be identified as the
wave-function correction (superscript wv), the term depending on $\Pi_\phi^{\cal A}$ as
the correction from $2\leftrightarrow 2$ scatterings and $1\leftrightarrow 3$
decays and inverse decays, that we collectively denote as scatterings
(superscript sca), such that we may write
\begin{align}
{\cal B}={\cal B}^{\rm wv}+{\cal B}^{\rm sca}\,.
\end{align}
For the following calculations, we choose $M=p^0>0$, for definiteness, and
note that ${\cal B}$ should be odd in $p^0$.
The wave-function correction is given by
\begin{align}
{\cal B}^{\rm wv}
=&g_w Y^2\int\frac{d^3 k}{(2\pi)^3} \sum\limits_{k^0=-p^0\pm|\mathbf k|}
p^02\pi\sum\limits_{\pm^\prime}\delta(-p^0\pm|\mathbf k|\pm^\prime|\mathbf k|)
\frac{1}{4{k^0}^2}
\\\notag
\times&
\left(
\frac{\partial}{\partial k^0}-\frac{1}{k^0}
\right)
\Pi_\phi^{H}(k)
\left[f_\ell((p+k)\cdot u)+f_\phi(k \cdot u)\right]
\,,
\end{align}
where we have made use of Eq.~(\ref{ret:distr}).
Clearly, we obtain only contributions for $\pm=\pm^\prime=+$
when $p^0>0$.
According to our decomposition above, this can be written as a
sum of a term depending on $\Pi^{{H},{\rm vac}}$ and
$\Pi^{{H},T\not=0}$,
\begin{align}
{\cal B}^{\rm wv}
={\cal B}^{{\rm wv},{\rm vac}}+{\cal B}^{{\rm wv},T\not=0}\,.
\end{align}
The contribution ${\cal B}^{{\rm wv},T\not=0}$ can be evaluated 
numerically, while
${\cal B}^{{\rm wv},{\rm vac}}$ inherits the IR
divergence from $\Pi_\phi^{{H},{\rm vac}}$, Eq.~(\ref{PiHvac}).
We can cast it to the
form ($|\mathbf p|=0$)
\begin{align}
\label{B:virt:vac}
{\cal B}^{{\rm wv},{\rm vac}}
=-\frac{g_w Y^2G M^2}{2^7\pi^4}\log\frac{\bar\Lambda}{\lambda}
\int d\Omega
\left[
f_{\ell}((p+k)\cdot u)+f_{\phi}(k\cdot u)
\right]
\,.
\end{align}

The scattering correction is
\begin{align}
{\cal B}^{\rm sca}
=& g_w Y^2 \int\frac{d^3 k}{(2\pi)^3}
\sum\limits_{k^0=-p^0\pm|\mathbf k|}
\frac{2 p^0}{(k^2)^2}
\Pi^{\cal A}(k)\left[f_\ell( (p+k)\cdot u )+f_\phi(k\cdot u)\right]\,.
\end{align}
Following the splitting of $\Pi_\phi^{\cal A}$ into vacuum and $T\not=0$ contributions,
we decompose as well
\begin{align}
{\cal B}^{\rm sca}={\cal B}^{{\rm sca},{\rm vac}}+{\cal B}^{{\rm sca},T\not=0}
\,.
\end{align}
First, we show that  ${\cal B}^{{\rm sca},T\not=0}$ yields a finite result
when $\lambda\to 0$. For this purpose, note that for small
$|k^0-|\mathbf k||$, one may expand for $k^2>0$
\begin{subequations}
\label{PiA:Tnot=0:expanded}
\begin{align}
\Pi_\phi^{{\cal A},{T\not=0}}(k)
=&\frac{G}{8\pi}\frac{k^2}{|\mathbf k|}
\Bigg[
2|\mathbf k|
-\frac2\beta\log\left|1-{\rm e}^{\beta\frac{k^0+|\mathbf k|}{2}}\right|
+\frac2\beta\log\left(\beta\frac{k^0-|\mathbf k|}{2}\right)
+\frac{k^0-|\mathbf k|}{2}+\cdots
\Bigg]
\end{align}
and for $k^2<0$
\begin{align}
\Pi_\phi^{{\cal A},{T\not=0}}(k)
=&\frac{G}{8\pi}\frac{k^2}{|\mathbf k|}
\Bigg[
2k^0
-\frac2\beta\log\left|1-{\rm e}^{\beta\frac{k^0+|\mathbf k|}{2}}\right|
+\frac2\beta\log\left(\beta\frac{|\mathbf k|-k^0}{2}\right)
+\frac{|\mathbf k|-k^0}{2}+\cdots
\Bigg]
\,.
\end{align}
\end{subequations}
Therefore, within ${\cal B}^{{\rm sca},T\not=0}$, we can integrate over the singularity
at $k^2=0$ in the principal value sense.

Next, we demonstrate that ${\cal B}^{{\rm sca},{\rm vac}}$ depends on $\lambda$ in
a way that cancels the $\lambda$ dependence within ${\cal B}^{{\rm wv},{\rm vac}}$.
For this purpose, we evaluate ($\mathbf p=\mathbf 0$)
\begin{align}
{\cal B}^{{\rm sca},{\rm vac}}=&
-g_w Y^2\int\frac{d^3k}{(2\pi)^3}\sum\limits_{k^0=-p^0\pm|\mathbf k|}
\frac{2}{p^0 \mp 2|\mathbf k|}
\frac{G}{8\pi}
\vartheta({p^0}^2\mp2p^0 |\mathbf k|-\lambda^2){\rm sign}k^0
\\\notag\times&
\left[
f_{\ell}((p+k)\cdot u)+f_{\phi}(k\cdot u)
\right]
=:{\cal B}^{{\rm sca},{\rm vac}+}+{\cal B}^{{\rm sca},{\rm vac}-}
\,.
\end{align}
While for ${\cal B}^{{\rm sca},{\rm vac}-}$, we can set $\lambda=0$ and
perform the integral numerically, we proceed with ${\cal B}^{{\rm sca},{\rm vac}+}$
as
\begin{align}
\label{B:real:vac+}
{\cal B}^{{\rm sca},{\rm vac}+}=&
g_wY^2\frac{G}{4\pi}
\int d\Omega
\int\limits_0^{\frac12(M-\lambda^2/M)}
\frac{\mathbf k^2 d|\mathbf k|}{(2\pi)^3}
\frac{1}{M-2|\mathbf k|}
\left[
f_{\ell}((p+k)\cdot u)+f_{\phi}(k\cdot u)
\right]
\\\notag
=&
\frac{g_w Y^2 G}{8\pi}\int\frac{d\Omega}{(2\pi)^3}
\Bigg\{
\\\notag
-&\left[
|\mathbf k|^2 \log(M-2|\mathbf k|)
\left(
f_{\ell}((p+k)\cdot u)+f_{\phi}(k\cdot u)
\right)
\right]^{|\mathbf k|=\frac12(M-\lambda^2/M)}_{|\mathbf k|=0}
\\\notag
+&
\int\limits_0^{M/2}
d|\mathbf k|
\log(M-2|\mathbf k|)
\frac{\partial}{\partial|\mathbf k|}
\mathbf k^2
\left(
f_{\ell}((p+k)\cdot u)+f_{\phi}(k\cdot u)
\right)
\Bigg\}
\,.
\end{align}
Therefore, the terms $\propto \log\lambda^2$ cancel between ${\cal B}^{\rm wv}$
and ${\cal B}^{\rm sca}$.
We emphasise that this cancellation works out, because
$\Pi_\phi^{{\cal A},{\rm vac}}$ and $\Pi_\phi^{H,{\rm vac}}$ are imaginary and
real part of the same analytic self-energy, which implies
relation~(\ref{rel:spec:herm}).

While these expressions are already defined within the text above, we finally
collect the explicit expressions for the various IR-finite contributions:
\begin{align}
{\cal B}^{{\rm sca,vac}-}
=&g_w Y^2\int\frac{d^3 k}{(2\pi)^3}\frac{1}{M+2|\mathbf k|}\frac{G}{4\pi}
\left[
f_\ell((p+k)\cdot u)+f_\phi(k\cdot u)
\right]
\,,
\end{align}
where $k^0=-M-|\mathbf k|$\,,
\begin{align}
{\cal B}^{{\rm sca},T\not=0}
=&g_w Y^2 \int\frac{d^3k}{(2\pi)^3}\sum\limits_{k^0=-M\pm|\mathbf k|}
\frac{2 M}{(k^2)^2}\Pi^{{\cal A},T\not=0}(k)
\left[
f_\ell((p+k)\cdot u)+f_\phi(k\cdot u)
\right]
\end{align}
with $\Pi^{{\cal A},T\not=0}$ given by Eqs.~(\ref{Pi:spectral:phi})
and
\begin{align}
{\cal B}^{{\rm wv},T\not=0}
=&g_wY^2\int\frac{d^3 k}{(2\pi)^3}2\pi\delta(M-2|\mathbf k|)
\frac{G T^2}{4 M^2}
\left(1+\frac M2\frac{d}{dk^0}\right)
\left[
f_\ell((p+k)\cdot u)+f_\phi(k\cdot u)
\right]\,,
\end{align}
where $k^0=-M+|\mathbf k|=-M/2$.
In the last expression, we have substituted $\Pi^{H,T\not=0}(k)$,
Eq.~(\ref{PiH:Tnot=0}) and have made use of the fact that
the non-HTL part of $\Pi^{H,T\not=0}(k)$
[{\it i.e.} the integral term
in Eq.~(\ref{PiH:Tnot=0})] and its derivative with respect to
$k^0$ vanish for $k^2=0$. We also note that ${\cal B}^{{\rm wv},T\not=0}$
corresponds to the phase-space suppression due to the asymptotic thermal Higgs boson mass
(note the minus sign from the statistical factors in the square brackets, as
$k^0<0$).
Together with the IR-divergent pieces,
these expressions above can be used in order to calculate the the
NLO wave-function corrections to the right-handed neutrino production
rate summarised in Eq.~(\ref{result:wv}).

\subsection{Radiation from the Fermion}
\label{sec:rad:fermion}

We now adapt the same strategy as for radiation from the scalar propagator
to radiation from the fermionic one. As a complication,
besides the cancellation between scattering and wave-function corrections
from vacuum loops,
there is also a cancellation between scattering and wave-function hard thermal loop (HTL)
contributions.

We express the leptonic self-energies as
$\slashed\Sigma_\ell=P_{\rm R}\gamma^\mu\Sigma_{\ell\mu}$. In the approximation
of massless particles in the loop, the spectral self-energy
is given by
\begin{subequations}
\label{Sig:A}
\begin{align}
\Sigma^{{\cal A}0}_\ell(k)
&=\frac{G T^2}{8\pi|\mathbf k|}
I_1\left({\frac{k^0}{T},\frac{|\mathbf k|}{T}}\right)\,,\\
\Sigma^{{\cal A}i}_\ell(k)
&=\frac{G T^2}{8\pi|\mathbf k|}
\left[
\frac{k^0}{|\mathbf k|}
I_1\left({\frac{k^0}{T},\frac{|\mathbf k|}{T}}\right)
-\frac{(k^0)^2-\mathbf k^2}{2|\mathbf k|T}
I_0\left({\frac{k^0}{T},\frac{|\mathbf k|}{T}}\right)
\right]\frac{k^i}{|\mathbf k|}\,,
\end{align}
\end{subequations}
where
\begin{subequations}
\label{I_01}
\begin{align}
I_0(y^0,y)=&I_0^{\rm vac}(y^0,y)+I_0^{T\not=0}(y^0,y)\,,
\\
I_0^{T\not=0}(y^0,y)=&-\vartheta(y^2-(y^0)^2)y^0-y
-\vartheta((y^0)^2-y^2)y\,{\rm sign}(y^0)
\\\notag
+&\log
\left|
\frac{1+{\rm e}^{\frac12(y^0+y)}}{1+{\rm e}^{\frac12(y^0-y)}}
\right|
+\log
\left|
\frac{1-{\rm e}^{\frac12(y^0+y)}}{1-{\rm e}^{\frac12(y^0-y)}}
\right|\,,
\\
I_0^{\rm vac}(y^0,y)=&\vartheta((y^0)^2-y^2)y{\rm sign}(y^0)\,,
\\
I_1(y^0,y)=&I_1^{\rm vac}(y^0,y)+\bar I_1^{T\not=0}(y^0,y)+I_1^{\rm HTL}(y^0,y)\,,
\\
\bar I_1^{T\not=0}(y^0,y)=&-\vartheta(y^2-(y^0)^2)\frac{(y^0)^2}{2}
-\frac12\vartheta((y^0)^2-y^2)|y^0|y
\\\notag
+&{\rm Re}\left[
x(\log(1+{\rm e}^x)-\log(1-{\rm e}^{x-y^0}))
+{\rm Li}_2(-{\rm e}^{x})-{\rm Li}_2({\rm e}^{x-y^0})
\right]^{x=\frac12(y^0+y)}_{x=\frac12(y^0-y)}\,,
\\
I_1^{\rm HTL}(y^0,y)=&\vartheta(y^2-(y^0)^2)\frac{\pi^2}{2}\,,
\\
I_1^{\rm vac}(y^0,y)=&\frac12\vartheta((y^0)^2-y^2)|y^0|y\,.
\end{align}
\end{subequations}
The bar on $\bar I_1^{T\not=0}(y^0,y)$ indicates that from this finite temperature 
term, the HTL contribution, which is purely thermal as well, is subtracted.

To this end, we also need the HTL-type contribution to the hermitian self-energy
\begin{subequations}
\begin{align}
\Sigma_\ell^{{H},{\rm HTL} 0}=&\frac{G T^2}{16|\mathbf k|}
\log\left|\frac{k^0+|\mathbf k|}{k^0-|\mathbf k|}\right|\,,
\\
\Sigma_\ell^{{H},{\rm HTL} i}=&\frac{GT^2k^0 k^i}{16|\mathbf k|^3}
\log\left|\frac{k^0+|\mathbf k|}{k^0-|\mathbf k|}\right|
-\frac{G T^2 k^i}{8|\mathbf k|^2}\,,
\end{align}
\end{subequations}
and the vacuum contribution
\begin{align}
\label{SigmaH:vac}
\slashed\Sigma_\ell^{{H},{\rm vac}}(k)=\frac{G}{8\pi^2}P_{\rm R}\slashed k\log\frac{\lambda}{\Lambda}
+\slashed k\delta Z_\ell
\,,
\end{align}
or, in dimensional regularisation,
\begin{align}
\slashed\Sigma_\ell^{{H},{\rm vac}}(k)=\frac{G}{8\pi^2}P_{\rm R}\slashed k\left(\log\frac{\lambda}{\mu}-\frac 12\Delta_\epsilon\right)
+\slashed k\delta Z_\ell
\,.
\end{align}
The term $\delta Z_\ell$ is a field-strength renormalisation that
cancels the divergences as $\Lambda\to \infty$ or $\epsilon\to 0$,
but is scheme dependent otherwise. In the following, we again implement its 
effect by replacing $\Lambda\to\bar\Lambda$, where $\bar\Lambda$
is finite. The non-HTL contribution to
the finite-temperature hermitian self-energy
$\slashed\Sigma^{H,T\not=0}$
is IR finite
and can be expressed in terms of a one-dimensional integral, a classic
result that can be found in Ref.~\cite{Weldon:1982bn}.

The leading correction to the fermionic spectral
function is
\begin{align}
S_\ell^{(1){\cal A}}&=
-\frac12\left(
{\rm i}S_\ell^{(0){\rm R}}{\rm i}\slashed\Sigma_\ell^{\rm R}{\rm i}S_\ell^{(0){\rm R}}
-{\rm i}S_\ell^{(0){\rm A}}{\rm i}\slashed\Sigma_\ell^{\rm A}{\rm i}S_\ell^{(0){\rm A}}
\right)
\\\notag
&=\frac{1}{2{\rm i}}
\left(
{\rm i}S_\ell^{(0){\rm R}}\slashed\Sigma_\ell^{H}{\rm i}S_\ell^{(0){\rm R}}
-{\rm i}S_\ell^{(0){\rm A}}\slashed\Sigma_\ell^{H}{\rm i}S_\ell^{(0){\rm A}}
\right)
\\\notag
&-\frac{1}{2}
\left(
{\rm i}S_\ell^{(0){\rm R}}\slashed\Sigma_\ell^{\cal A}{\rm i}S_\ell^{(0){\rm R}}
+{\rm i}S_\ell^{(0){\rm A}}\slashed\Sigma_\ell^{\cal A}{\rm i}S_\ell^{(0){\rm A}}
\right)
\,,
\end{align}
and the  corrections to the Wightman functions are
\begin{subequations}
\begin{align}
{\rm i}S_\ell^{(1)<}&=2S_\ell^{(1){\cal A}} (-f_\ell)\,,
\\
{\rm i}S_\ell^{(1)>}&=2S_\ell^{(1){\cal A}} (1-f_\ell)\,.
\end{align}
\end{subequations}
In the following, we attach the same superscripts as for the functions $I_{1,2}$
to various quantities, in order to indicate which loop terms they originate from.

The correction to the neutrino relaxation rate follows from Eq.~(\ref{calF})
\begin{align}
\label{I:fermi:rad}
{\cal F}={\rm tr}\left[\slashed p \slashed\Sigma_\ell^{\cal A}(p)\right]
=&g_w Y^2\int\frac{d^4 k}{(2\pi)^4}{\rm tr}
\left[
\slashed p {\rm i}S_\ell^{(1)>}(k){\rm i}\Delta^>_\phi(p-k)
-\slashed p {\rm i}S_\ell^{(1)<}(k){\rm i}\Delta^<_\phi(p-k)
\right]
\\\notag
=&g_w Y^2\int\frac{d^3 k}{(2\pi)^3}\sum\limits_{k^0=p^0\pm|\mathbf k|}
\frac{\mp 1}{2|p^0-k^0|}
{\rm tr}\left[\slashed p 2 S_\ell^{(1)\cal A}(k)\right]
\\\notag
\times&
\left[1-f_{\ell}(k\cdot u)+f_{\phi}((p-k)\cdot u)\right]
\\\notag
=:&{\cal F}^{\rm vac,fin}+{\cal F}^{\rm vac,col}+{\cal F}^{{\rm HTL},{\rm fin}}+{\cal F}^{{\rm HTL},{\rm col}}+{\cal F}^{T\not=0}
\,,
\end{align}
where the particular contributions to ${\cal F}$ are defined below.

The Dirac trace is evaluated as
\begin{align}
\label{Dirac:Trace}
{\rm tr}[\slashed p S_\ell^{(1){\cal A}}(k)]
=&
\left[
\left(\frac{\rm i}{k^2+{\rm i}\,{\rm sign} k^0 \varepsilon}\right)^2
-\left(\frac{\rm i}{k^2-{\rm i}\,{\rm sign} k^0 \varepsilon}\right)^2
\right]
\\\notag
\times&
\left[-(p^2+k^2)k\cdot{\rm i}\Sigma_\ell^{H}(k)+ k^2 p \cdot{\rm i}\Sigma_\ell^{H}(k) \right]
\\\notag
+&
\left[
\left(\frac{\rm i}{k^2+{\rm i}\,{\rm sign} k^0 \varepsilon}\right)^2
+\left(\frac{\rm i}{k^2-{\rm i}\,{\rm sign} k^0 \varepsilon}\right)^2
\right]
\\\notag
\times&
\left[-(p^2+k^2)k\cdot\Sigma_\ell^{\cal A}(k)+ k^2 p \cdot\Sigma_\ell^{\cal A}(k) \right]
\,.
\end{align}

As a consequence of Lorentz invariance,
$\Sigma^{{\cal A},{H},{\rm vac}\mu}(k)\propto k^\mu$,
which may be verified explicitly by inspection of Eqs.~(\ref{Sig:A},\ref{I_01},\ref{SigmaH:vac}). Therefore,
we define
\begin{align}
\slashed\Sigma_\ell^{{\cal A},{H},{\rm vac}}(k)
=P_{\rm R}\slashed k\hat\Sigma_\ell^{{\cal A},{H},{\rm vac}}(k)\,,
\end{align}
such that
\begin{subequations}
\begin{align}
\hat\Sigma_\ell^{{\cal A},{\rm vac}}(k)&=\frac{G}{16\pi}\vartheta(k^2-\lambda^2)
{\rm sign}(k^0)\,,
\\\notag
\hat\Sigma_\ell^{{H},{\rm vac}}(k)&=\frac{G}{8\pi^2}\log\frac{\lambda}{\bar \Lambda}
\,.
\end{align}
\end{subequations}
We have included here the threshold condition form the gauge-boson mass $\lambda$, that
is needed for the infrared regularisation.
We obtain
\begin{align}
\left[-4(p^2+k^2)k\cdot\Sigma_\ell^{{\cal A},{H},{\rm vac}}(k)+4 k^2 p \cdot\Sigma_\ell^{{\cal A},{H},{\rm vac}}(k) \right]
=-2(k^2 p^2+ (k^2)^2)\hat\Sigma_\ell^{{\cal A},{H},{\rm vac}}(k)\,.
\end{align}
The term $\propto (k^2)^2$ gives a finite contribution to
${\cal F}^{{\rm vac},{\rm fin}}$ and can be evaluated
straightforwardly.  The term $\propto p^2k^2$ gives rise to
${\cal F}^{{\rm vac},{\rm col}}={\cal F}^{{\rm vac},{\rm col},{\rm wv}}+{\cal F}^{{\rm vac},{\rm col},{\rm sca}}$. The
wave-function contribution ${\cal F}^{{\rm vac},{\rm col},{\rm wv}}$ originates from
$\Sigma_\ell^{{H},{\rm vac}}$ and the scattering
correction ${\cal F}^{{\rm vac},{\rm col},{\rm sca}}$
from $\Sigma_\ell^{{\cal A},{\rm vac}}$. These terms individually contain collinear divergences
that are regulated by the gauge-boson mass $\lambda$ and that cancel when added to
${\cal F}^{{\rm vac},{\rm col}}$, as we demonstrate now. We find
\begin{align}
\label{F:col:wv:vac}
{\cal F}^{{\rm vac},{\rm col},{\rm wv}}
=&
g_w Y^2\int\frac{d^3k}{(2\pi)^3}\frac{1}{2|\mathbf k|}2\pi\delta(p^0-2|\mathbf k|)
\frac{1}{4{k^0}^2}\frac{\partial}{\partial k^0}\frac{G}{8\pi^2}
p^2 k^2 \log\frac{\lambda}{\bar \Lambda}
\\\notag
\times&
\left[
1-f_{\ell}(k\cdot u)+f_{\phi}((p-k)\cdot u)
\right]
\\\notag
=&-\frac{g_w Y^2 G M^2}{2^8\pi^4}\log\frac{\bar\Lambda}{\lambda}
\int d\Omega
\left[
1-f_{\ell}(k\cdot u)+f_{\phi}((p-k)\cdot u)
\right]
\,,
\end{align}
where $k^0=p^0-|\mathbf k|$.
The scattering corrections yield
\begin{align}
{\cal F}^{{\rm vac},{\rm col},{\rm sca}}=&
g_w Y^2\int\frac{d^3k}{(2\pi)^3}\frac{1}{2|\mathbf k|}\sum\limits_{k^0=p^0\pm|\mathbf k|}\mp \frac{2 p^2}{k^2}
\frac{G}{16\pi} \vartheta(k^2-\lambda^2){\rm sign}k^0
\\\notag
\times&
\left[
1-f_{\ell}(k\cdot u)+f_{\phi}((p-k)\cdot u)
\right]
\\\notag
=:&
{\cal F}^{{\rm vac},{\rm col},{\rm sca}+}+{\cal F}^{{\rm vac},{\rm col},{\rm sca}-}
\,.
\end{align}
Again, we analytically isolate the logarithmic dependence on $\lambda$, which
is for the $-$ contribution
\begin{align}
\label{F:col:sca:vac}
{\cal F}^{{\rm vac},{\rm col},{\rm sca}-}=&
\frac{g_w Y^2 G}{128\pi^4}\int d\Omega\int\limits_0^{\frac{p^0}{2}-\frac{\lambda^2}{2p^0}}
|\mathbf k| d|\mathbf k|\frac{p^0}{p^0-2|\mathbf k|}\left[1-f_{\ell}(k\cdot u)+f_{\phi}((p-k)\cdot u)\right]
\\\notag
=&
-\frac{g_w Y^2 G}{2^8\pi^4}\int d\Omega\Bigg\{
\big[\log (M-2|\mathbf k|)M|\mathbf k|
\\\notag
\times&
\left[1-f_{\ell}(k\cdot u)+f_{\phi}((p-k)\cdot u)\right]\big]^{|\mathbf k|=\frac{M}{2}-\frac{\lambda^2}{2 M}}_{|\mathbf k|=0}
\\\notag
+&\int\limits_0^{\frac{M}{2}}d|\mathbf k|
\log(M-2|\mathbf k|)\frac{\partial}{\partial|\mathbf k|} M|\mathbf k|\left[1-f_{\ell}(k\cdot u)+f_{\phi}((p-k)\cdot u)\right]
\Bigg\}\,.
\end{align}
The logarithmic dependence on $\lambda$ is therefore cancelled cancelled with
${\cal F}^{{\rm vac},{\rm col},{\rm wv}}$, Eq.~(\ref{F:col:wv:vac}). This is
again a consequence of the fact that $\slashed\Sigma^{{\cal A},H,{\rm vac}}$
are the anti hermitian and hermitian parts of the same analytic self energy,
evaluated at the two-particle branch cut.

The HTL-type contributions are decomposed into terms originating from 
$\slashed\Sigma_{\ell}^{H}$ and $\slashed\Sigma_{\ell}^{\cal A}$ as well,
${\cal F}^{\rm HTL}={\cal F}^{{\rm HTL},{\rm wv}}+{\cal F}^{{\rm HTL},{\rm sca}}$.
We observe that
$k\cdot\Sigma^{{\cal A},{\rm HTL}}(k)=0$ and
$k\cdot\Sigma^{{H},{\rm HTL}}(k)$ is finite for $k^2=0$, whereas
$p\cdot\Sigma^{{H},{\rm HTL}}(k)$ has a logarithmic divergence
for $k^2\to0$.
The terms that are $\propto p\cdot\Sigma^{{H},{\rm HTL}}(k)$
and $\propto p\cdot\Sigma^{{\cal A},{\rm HTL}}(k)$
in Eq.~(\ref{Dirac:Trace}) therefore lead to collinearly divergent contributions
to ${\cal F}^{\rm HTL}$, and we denote these by
${\cal F}^{{\rm HTL},{\rm col},{\rm wv}}$ and
${\cal F}^{{\rm HTL},{\rm col},{\rm sca}}$, whereas the terms
$\propto k\cdot\Sigma^{{H},{\rm HTL}}(k)$
and $\propto k\cdot\Sigma^{{\cal A},{\rm HTL}}(k)$ are finite and referred to
as ${\cal F}^{{\rm HTL},{\rm fin},{\rm wv}}$ and
${\cal F}^{{\rm HTL},{\rm fin},{\rm sca}}$, such that
\begin{subequations}
\begin{align}
{\cal F}^{{\rm HTL},{\rm col}}=&
{\cal F}^{{\rm HTL},{\rm col},{\rm wv}}
+{\cal F}^{{\rm HTL},{\rm col},{\rm sca}}\,,
\\
{\cal F}^{{\rm HTL},{\rm fin}}=&
{\cal F}^{{\rm HTL},{\rm fin},{\rm wv}}
+{\cal F}^{{\rm HTL},{\rm fin},{\rm sca}}\,.
\end{align}
\end{subequations}
The finite terms can be integrated in a straightforward manner, while the
collinearly divergent contributions must be regularised, with a cancellation
of the dependence on the regulator in the sum of wave-function and scattering
contributions.
In order to regulate these terms, we set the gauge-boson mass $\lambda=0$, but we
introduce a lepton mass $m_\ell$. Then, we find that
\begin{align}
\label{F:HTL:col:wv}
{\cal F}^{{\rm HTL},{\rm col},{\rm wv}}=&
-g_w Y^2\int\frac{d^3k}{(2\pi)^3}\frac{1}{|\mathbf k|}
2\pi\delta(p^0-|\mathbf k|-\sqrt{\mathbf k^2+m_\ell^2})
\\\notag
\times&
\frac{1}{4{k^0}^2}\frac{\partial}{\partial k^0}
k^2
p\cdot\Sigma_\ell^{{H},{\rm HTL}}(k)
\left[1-f_{\ell}(k\cdot u)+f_{\phi}((p-k)\cdot u)\right]
\\\notag
=&\frac{g_w Y^2 G T^2}{2^7\pi^2}\log\frac{m_\ell^2}{M^2}
\int d\Omega \left[1-f_{\ell}(k\cdot u)+f_{\phi}((p-k)\cdot u)\right]
\end{align}
and
\begin{align}
\label{F:HTL:col:sca}
{\cal F}^{{\rm HTL},{\rm col},{\rm sca}}=&
-g_w Y^2\int\frac{d^3k}{(2\pi)^3}\frac{1}{2|\mathbf k|}
\frac{4}{(k^2-m_\ell^2)^2}k^2 p\cdot\Sigma_\ell^{{\cal A},{\rm HTL}}(k)
\\\notag
\times&
\left[1-f_{\ell}(k\cdot u)+f_{\phi}((p-k)\cdot u)\right]
\\\notag
&\hskip-1.5cm =
\frac{g_w Y^2 G T^2}{2^7\pi^2}\int d\Omega
\Bigg\{
\big[\log\left(m_\ell^2+(2|\mathbf k|-M)M\right)
\\\notag
\times&
\left[1-f_\ell(k\cdot u)+f_\phi((p-k)\cdot u)\big]\right]^{|\mathbf k|=\infty}_{|\mathbf k|=\frac{M}{2}}
\\\notag
&\hskip-1.5cm -
\int\limits_{\frac{M}{2}}^\infty d|\mathbf k|\log\left(m_\ell^2+(2|\mathbf k|-M)M\right)\frac{\partial}{\partial|\mathbf k|}
\left[1-f_{\ell}(k\cdot u)+f_{\phi}((p-k)\cdot u)\right]
\Bigg\}\,.
\end{align}
Notice that there is no $+$ contribution from the sum over $\pm$ in
Eq.~(\ref{I:fermi:rad}), because $\Sigma^{{\rm HTL}{\cal A}}(k)$ vanishes for
positive $k^0$.
The logarithmic dependence on $m_\ell^2$ therefore cancels when adding scattering
and wave-function
corrections, Eqs.~(\ref{F:HTL:col:sca}) and~(\ref{F:HTL:col:wv}). Once
more, the prefactors for the cancellation match
because $\Sigma^{{\rm HTL}{\cal A},H}(k)$
are anti hermitian and hermitian part of an analytic self-energy evaluated at
the branch cut.
Notice, that the HTL-type contributions diverge logarithmically when
$p^2\to 0$. In that situation, one should use the resummed lepton
propagator~\cite{Arnold:2000dr,Arnold:2001ms,Besak:2012qm,GSG}.

The result for ${\cal F}^{T\not=0}$ [that is defined as the contribution
to $\cal F$ from $\slashed\Sigma^{H,T\not=0}$ and from $\slashed\Sigma^{{\cal A},T\not=0}$,
which in turn results from replacing
$I_0\to I_0^{T\not=0}$ and $I_1\to \bar I_1^{T\not=0}$ in Eq.~(\ref{Sig:A})]
can be obtained by integrating
over the pole at $k^2=0$ in the principal value sense, what can be verified by
checking the limiting behaviour of $I_{0}^{T\not=0}(y^0,y)$ and
$\bar I_{1}^{T\not=0}(y^0,y)$ for
$y^0 \to y$, {\it i.e.} that these functions are continuous at that point.

For completeness, we again list the explicit expressions for the various
IR-finite terms:
\begin{align}
{\cal F}^{\rm vac,fin,wv}=0\,,\hskip6cm
\end{align}
\begin{align}
{\cal F}^{\rm vac,fin,sca}
=&g_w Y^2 \int\frac{d^3 k}{(2\pi)^3}\frac{1}{2|\mathbf k|}
\sum\limits_{k^0=M\pm|\mathbf k|}
\mp \frac{G}{8\pi}\vartheta(k^2)\,{\rm sign}(k^0)
\\\notag
\times&
\left[1-f_\ell(k\cdot u)+f_\phi((p-k)\cdot u)\right]
\,,
\end{align}
\begin{align}
{\cal F}^{{\rm vac,col,sca}+}
=&-g_w Y^2 \int\frac{d^3 k}{(2\pi)^3}\frac{1}{2|\mathbf k|}
\frac{M}{M+2|\mathbf k|}\frac{G}{8\pi}
\left[1-f_\ell(k\cdot u)+f_\phi((p-k)\cdot u)\right]_{k^0=M+|\mathbf k|}
\,,
\end{align}
\begin{align}
{\cal F}^{\rm HTL,fin,wv}=&-g_w Y^2\int\frac{d^3k}{(2\pi)^3}\frac{1}{|\mathbf k|}
2\pi\delta(M-2|\mathbf k|)
\left(-\frac{1}{4 {k^0}^2}\frac{d}{dk^0}+\frac{1}{4{k^0}^3}\right)
\\\notag
\times&
\left(M^2+{k^0}^2-\mathbf k^2\right)
\frac{GT^2}{8}\left[1-f_\ell(k\cdot u)+f_\phi((p-k)\cdot u)\right]\,,
\end{align}
where $k^0=\frac M2$,
\begin{align}
{\cal F}^{\rm HTL,fin,sca}
=&0\,,\hskip6cm
\end{align}
\begin{align}
{\cal F}^{T\not=0}
=&g_w Y^2 \int\frac{d^3 k}{(2\pi)^3}\frac{1}{2|\mathbf k|}
\Bigg\{
\\\notag
&\sum\limits_{k^0=M\pm|\mathbf k|}
\frac{\mp 1}{(k^2)^2}
\left[
2(p^2+k^2) k\cdot\Sigma_\ell^{{\cal A},T\not=0}(k)
-2k^2 p\cdot \Sigma_\ell^{{\cal A},T\not=0}(k)
\right]
\\\notag
\times&
\left[1-f_\ell(k\cdot u)+f_\phi((p-k)\cdot u)\right]
\\\notag
+&2\pi\delta(M-2|\mathbf k|)
\Big[
\left(\frac{1}{2{k^0}^2}\frac{d}{d k^0}-\frac{1}{2{k^0}^3}\right)\left((p^2+k^2) k\cdot\Sigma_\ell^{H,T\not=0}(k)
-k^2 p\cdot \Sigma_\ell^{H,T\not=0}(k)\right)
\\\notag
\times&
\left[1-f_\ell(k\cdot u)+f_\phi((p-k)\cdot u)\right]
\Big]_{k^0=\frac M2}
\Bigg\}\,,
\end{align}
where ${\cal F}^{\rm vac,fin}={\cal F}^{\rm vac,fin,wv}+{\cal F}^{\rm vac,fin,sca}$.
These expressions can be substituted into Eq.~(\ref{result:wv}), in order
to obtain the production rate for right-handed neutrinos. Notice also that the sign
of ${\cal F}^{\rm HTL,fin,wv}$ is consistent with the phase-space suppression due
to the asymptotic thermal lepton mass.

\section{Vertex Contribution}
\label{section:VERT}

\subsection{Scatterings or Real Emissions}
\label{sec:vertex:real}

The $2\leftrightarrow 2$ and $1\leftrightarrow 3$ contributions (involving
the real emission of gauge bosons, here collectively referred to
as scatterings) to the thermal
production rate of $N$ follow from the first term of
Eq.~(\ref{Sigmaless:start}),
\begin{align}
\label{Sigmaless:sca}
{\rm tr}[\slashed p  {\rm i}\slashed \Sigma_N^<&(p)]^{\rm sca}
=4 g_w G Y^2\int\frac{d^4 k}{(2\pi)^4}\frac{d^4 q}{(2\pi)^4}
[2(k-p)+q]^\nu{\rm tr}\Big[\slashed p
\\\notag
\times&{\rm i}S_\ell^H(k)\gamma^\mu
{\rm i}S_\ell^{<}(k+q){\rm i}\Delta_\phi^{<}(-k+p)
{\rm i}\Delta_{\mu\nu}^>(q){\rm i}\Delta^H_\phi(-k+p-q)
\,.
\end{align}
In order to bring the integrand to a symmetric form,
we replace the momenta
\begin{align}
q\to -r\,,\qquad k\to \frac p2-q+\frac r2\,,
\end{align}
which takes the effect
\begin{align}
-k+p&\to \frac p2+q-\frac r2\,,\qquad k+q \to \frac p2-q-\frac r2\,,\qquad -k+p-q \to \frac p2+q+\frac r2\,.
\end{align}
Furthermore, the on-shell delta-functions of the Wightman propagators force
\begin{subequations}
\begin{align}
\left(\frac p2+q-\frac r2\right)^2=&0\,,\\
\left(\frac p2-q-\frac r2\right)^2=&0\,,\\
r^2=&\lambda^2\,,
\end{align}
\end{subequations}
where $\lambda$ again is a fictitious gauge-boson mass,
such that this contribution reads
\begin{align}
{\rm tr}[\slashed p  {\rm i}\slashed \Sigma_N^<(p)]^{\rm sca}
=&4 g_w G Y^2\int\frac{d^4 q}{(2\pi)^4}\frac{d^4 r}{(2\pi)^4}
\\\notag
\times&
\bigg(p^2\left[-2p^2-2\lambda^2+2\left(\frac p2+q+\frac r2\right)^2+\left(\frac p2-q+\frac r2\right)^2\right]
\\\notag
+&\left[\frac p2+q+\frac r2\right]^2\left[\frac p2-q + \frac r2\right]^2\bigg)
{\rm PV}\left[\frac{1}{\left(\frac p2-q+\frac r2\right)^2}\frac{1}{\left(\frac p2+q+\frac r2\right)^2}\right]
\\\notag
\times&
2\pi\delta\left(\left(\frac p2-q-\frac r2\right)^2\right)
2\pi\delta\left(\left(\frac p2 +q-\frac r2\right)^2\right)
2\pi\delta\left(r^2-\lambda^2\right)
\\\notag
\times&
{\rm sign}\left(\frac{M}{2}-q^0-\frac{r^0}{2}\right)
{\rm sign}\left(\frac{M}{2}+q^0-\frac{r^0}{2}\right)
{\rm sign}\left(r^0\right)
\\\notag
\times&
f_\ell\left(\left((\frac p2-q-\frac r2\right)\cdot u\right)
f_\phi\left(\left(\frac p2+q-\frac r2\right)\cdot u\right)
\left[1+f_A(-r\cdot u)\right]\,.
\end{align}
We integrate first over $dr^0$, $dq^0$ and $d\cos\vartheta$ using the three
on-shell delta-functions and then change the integration variables $|\mathbf r|$
and $|\mathbf q|$ back to $r^0$
and $q^0$, {\it i.e.} the integration measure changes as
\begin{align}
&\int\frac{d^4 q}{(2\pi)^4}\frac{d^4 r}{(2\pi)^4}
2\pi\delta\left(\left(\frac p2-q-\frac r2\right)^2\right)
2\pi\delta\left(\left(\frac p2 + q-\frac r2\right)^2\right)
2\pi\delta\left(r^2-\lambda^2\right)
\\\notag
\to&
\sum\limits_{\underset{q^0=\pm^\prime|q^0|}{r^0=\pm|r^0|}}
\frac{1}{(2\pi)^5}
\int\limits_0^{2\pi}d\varphi\int\limits_0^{2\pi}d\psi
\int\limits_{-1}^1 d\cos\varrho
\int\limits_{\cal{I}_\mathbf{q}}\mathbf{q}^2d|\mathbf q|
\int\limits_{\cal{I}_\mathbf{r}}\mathbf{r}^2d|\mathbf r|
\frac{1}{2\sqrt{\mathbf r^2+\lambda^2}}\frac{1}{4|q^0|}\frac{1}{|\mathbf q||\mathbf r|}
\\\notag
\to&
\frac 18
\frac{1}{(2\pi)^5}
\int\limits_0^{2\pi}d\varphi\int\limits_0^{2\pi}d\psi
\int\limits_{-1}^1 d\cos\varrho
\int\limits_{{\cal I}_{q^0}}d q^0
\int\limits_{{\cal I}_{r^0}}d r^0
\,,
\end{align}
where $\varrho$ denotes the angle between $\mathbf{q}$ and
$\mathbf{u}$. More specifically, we choose to parametrise the angular dependences as
\begin{align}
\mathbf r=|\mathbf r|\left(\begin{array}{c}\sin\vartheta\sin\varphi\\\sin\vartheta\cos\varphi\\\cos\vartheta\end{array}\right)
\,,\quad
\mathbf q=|\mathbf q|\left(\begin{array}{c}0\\0\\1\end{array}\right)
\,,\quad
\mathbf u=\frac{|\tilde{\mathbf{p}}|}{M}\left(\begin{array}{c}\sin\varrho\sin\psi\\ \sin\varrho\cos\psi\\ \cos\varrho\end{array}\right)
\,.
\end{align}

The integration domains ${\cal I}$ are best determined
from distinguishing between the various kinematic situations, which are
sketched in Figure~\ref{fig:kinregions}. Besides, we take within these
regions a positive integration measure, such that we do not obtain
explicit minus signs from the Jacobians. The regions follow from
the three on-shell constraints. Notice that these
conditions may be combined
to obtain
\begin{align}
\label{q:real}
|\mathbf q|=\frac 12\sqrt{M^2 + 4 {q^0}^2 -2 M r^0+\lambda^2}\,,
\end{align}
and when using this as a relation for $\cos\vartheta$,
\begin{align}
\label{rel:costheta}
\left(\frac p2+q-\frac r2\right)^2=\left(\frac{M}{2}+q^0\right)^2-\left(\frac{M}{2}+q^0\right)r^0+|\mathbf q||\mathbf r|\cos\vartheta+\frac{\lambda^2}{4}-\mathbf q^2=0.
\end{align}
The relation $-1\leq\cos\vartheta\leq1$ is then one of the constraints
that must be imposed in order
to determine the regions of integration in Figure~\ref{fig:kinregions}. Indeed, for $\lambda = 0$ this inequality reduces to:
\begin{align}
\left(r^0 - \frac{M}{2}\right)\left({q^0}^2-\frac{1}{4}{r^0}^2\right) \geq 0\,.
\end{align}

Besides, we note that the on-shell constraints imply that
the invariant momentum squares of the off-shell propagators are
\begin{subequations}
\begin{align}
\left(\frac p2+q+\frac r2\right)^2=&M\left(2 q^0 + r^0\right)\,,
\\
\left(\frac p2-q+\frac r2\right)^2=&M\left(-2 q^0 + r^0\right)\,.
\end{align}
\end{subequations}
This suggests a change of variables to
\begin{subequations}
\label{xy:real}
\begin{align}
x&=q^0 +\frac 12 r^0\,,
\\
y&=-q^0 +\frac 12 r^0\,,
\end{align}
\end{subequations}
which has the Jacobian $-1$. The integration regions in terms of these variables
are indicated in Figure~\ref{fig:kinregions} as well. The collinear divergences
are located on the fringes
of the integration region where either $x$ or $y$ vanish, soft divergences are at the
coincident point $x=y=0$.

We can therefore recast the scattering contributions to the
neutrino production as
\begin{align}
\label{N:production:real:xy}
{\rm tr}[\slashed p   {\rm i} \slashed\Sigma_N^<(p)]^{\rm sca}
=&4 g_w G Y^2
\frac{1}{32}
\frac{1}{(2\pi)^4}
\int\limits_0^{2\pi}d(\varphi-\psi)
\int\limits_{-1}^1 d\cos\varrho
\int\limits_{{\cal I}_{x}}d x
\int\limits_{{\cal I}_{y}}d y
\\\notag
\times&
{\rm sign}(x+y){\rm sign}\left(\frac{M}{2}-x\right){\rm sign}\left(\frac{M}{2}-y\right)
{\rm PV}\frac{-2 M^2 + 4M x +2 M y +4 xy}{xy}
\\\notag
\times&
f_\ell\left(\left(p/2-q-r/2\right)\cdot u\right)
f_\phi\left(\left(p/2 + q-r/2\right)\cdot u\right)
\left[1+f_A(-r\cdot u)\right]\,.
\end{align}
The statistical functions can be evaluated using above relations.
In particular, the scalar products that appear in the arguments are given by
\begin{subequations}
\begin{align}
p\cdot u=&\tilde p^0\,,
\\
q\cdot u=&
\frac12
(x-y)\frac{\tilde p^0}{M}
-|\mathbf q|\frac{|\tilde{\mathbf p}|}{M}\cos\varrho\,,
\\
r\cdot u=&(x+y)\frac{\tilde p^0}{M}
-|x+y|\frac{|\tilde{\mathbf p}|}{M}
\left[
\cos\varrho\cos\vartheta+\sin\varrho\sin\vartheta\cos(\varphi-\psi)
\right]\,.
\end{align}
\end{subequations}
with $|\mathbf q|$ as in Eq.~(\ref{q:real}) and $\cos\vartheta$ given by Eq.~(\ref{rel:costheta}).
Close to the collinear edges, where $x=0\lor y=0$, we can evaluate
\begin{subequations}
\label{scat:arguments}
\begin{align}
\left(\frac p2-q-\frac r2\right)\cdot u\approx&\left(\frac 12-\frac xM\right)(\tilde p^0+|\tilde{\mathbf{p}}|\cos\varrho\;{\rm sign}(M-x){\rm sign}(M-y))
\\\notag
+&\frac{\sqrt{xy}}{M}|\tilde{\mathbf{p}}|\sin\varrho\cos(\varphi-\psi)\,,
\\
\left(\frac p2+q-\frac r2\right)\cdot u\approx&\left(\frac 12-\frac yM\right)(\tilde p^0-|\tilde{\mathbf{p}}|\cos\varrho\;{\rm sign}(M-x){\rm sign}(M-y))
\\\notag
+&\frac{\sqrt{xy}}{M}|\tilde{\mathbf{p}}|\sin\varrho\cos(\varphi-\psi)\,,
\\
r\cdot u\approx&\frac{x+y}{M}\tilde p^0+\frac{x-y}{M}|\tilde{\mathbf{p}}|\cos\varrho\;{\rm sign}(M-x){\rm sign}(M-y)
\\\notag
-&\frac{2\sqrt{xy}}{M}|\tilde{\mathbf{p}}|\sin\varrho\cos(\varphi-\psi)
\,.
\end{align}
\end{subequations}
Note that Eqs.~(\ref{scat:arguments}) are understood as leading order expansions and hence the $\sim\sqrt{xy}$ corrections are applicable only when both $x \sim y \approx 0$.
As we aim for rearranging the various contributions to the vertex-type correction
into a manifestly finite integral, we have dropped here the dependence on
the regulating gauge-boson mass $\lambda$. We have also let $\cos\varrho\to {\rm sign}(M-x){\rm sign}(M-y)\cos\varrho$ for later convenience.

In order to arrange for an IR-finite integral, where the collinear divergences
on the fringes $x=0$ and $y=0$ cancel, it is useful
to define the integrand
\begin{align}
\label{J:sca}
{\cal J}^{\rm sca}(x,y)=& K^{\rm sca}(x,y)
f_\ell\left((p/2-q-r/2)\cdot u\right)
f_\phi\left((p/2+q-r/2)\cdot u\right)
\\\notag
\times&
\left[1+f_A(-r\cdot u)\right]
{\cal S}(x,y)
\\\notag
\underset{|x|\ll M \dot\lor |y|\ll M}{\approx}&
-K^{\rm sca}(x,y)
f_\ell\left(\left(1-\frac{2x}{M}\right)E_s\right)
f_\phi\left(\left(1 - \frac{2y}{M}\right)E_{-s}\right)
\\\notag
&\times
f_A\left(\frac{2x}{M}E_s + \frac{2y}{M}E_{-s}\right)
{\cal S}(x,y)
\,,
\end{align}
where $s \equiv {\rm sign}(M-x){\rm sign}(M-y)$,
\begin{align}
\label{E:pm}
E_\pm \equiv \frac 12 (\tilde p_0\pm|\tilde{\mathbf{p}}|\cos\varrho)\,,
\end{align}
and ${\cal S}(x,y)$ is a step function defining the support of the integral in accordance with Figure~\ref{fig:kinregions}:
\begin{align}
\label{real:support}
{\cal S}(x,y) =& \vartheta(x)\vartheta(y)\vartheta(M/2 -x-y) + \vartheta(-x)\vartheta(-y)
\\\notag
 +& \big(\vartheta(x)\vartheta(-y) + \vartheta(-x)\vartheta(y)\big)\vartheta\left(x+y-M/2\right)\,.
\end{align}
For later convenience, we have separated the kinematic part in the factor
\begin{align}
\label{K:real}
K^{\rm sca}(x,y)=&{\rm sign}(x+y)
{\rm sign}\left(\frac M2 -x\right){\rm sign}\left(\frac M2 -y\right)
{\rm PV}\frac{-2M^2 + 4 M x + 2 M y+4xy}{xy}\,.
\end{align}
In the approximate form for ${\cal J}^{\rm sca}$, we have made the restriction
to $|x|\ll M \,\dot\lor\, |y|\ll M$, where $\dot \lor$ denotes the exclusive or, which justifies dropping 
the $\sim\sqrt{xy}$ corrections. We are going to use this expression in order to show the
 cancellation of collinear divergences. For the soft divergence at $x=y=0$, there occurs a complication
due to the Bose divergence of the gauge boson distribution, and we will
provide a separate discussion.

\subsection{Vertex Correction with on-Shell Gauge Boson}
\label{sec:vertex:virt:1}

We now bring the vertex contribution with an on-shell gauge boson to a form
that can be matched with the contributions from scatterings.
It is useful to reparametrise in Eq.~(\ref{Sigmaless:start})
\begin{align}
\label{reparametrisation:vertex}
k\to p-q\,,\qquad q\to q-k\,,
\end{align}
such that the second term in
Eq.~(\ref{Sigmaless:start}) turns into
\begin{align}
{\rm tr}[\slashed p{\rm i}\slashed \Sigma^<(p)]^{\rm vert}_1
=&(3 g_2^2 +g_1^2) Y^2\int\frac{d^4 k}{(2\pi)^4}\frac{d^4 q}{(2\pi)^4}
[-k-q]^\nu{\rm tr}\Big[\slashed p
\\\notag
\times&{\rm i}S_\ell^H(p-q)\gamma^\mu
{\rm i}S_\ell^<(p-k)
{\rm i}\Delta_\phi^<(k)
{\rm i}\Delta_\phi^H(q)
{\rm i}\Delta^F_{\mu\nu}(q-k)
\Big]\,.
\end{align}
We perform the integrations over $dk^0$ and $dq^0$ making use of the
on-shell delta-functions. When we define
\begin{align}
k_\pm=(\pm|\mathbf k|,\mathbf k)\,,\qquad
q_{1\pm}=(k^0\pm\omega,\mathbf q)\,,\qquad
\omega=\sqrt{(\mathbf q-\mathbf k)^2+\lambda^2}\,,
\end{align}
we obtain
\begin{align}
\label{Sigma:virt1:omega}
{\rm tr}[\slashed p {\rm i}\slashed \Sigma^<(p)]^{\rm vert}_1
=&
4 g_w G Y^2\frac{1}{16(2\pi)^4}
\sum_\pm\int\limits_0^{2\pi}d(\varphi-\psi)\int\limits_{-1}^{1}d\cos\varrho
\int\limits_\lambda^\infty d\omega \int\limits_{q_{\rm min}}^{q_{\rm max}}
d|\mathbf q| \frac{|\mathbf q|}{M}
\\\notag
\times&
{\rm PV}\bigg[
\frac{2M^4}{\left[(M/2\mp \omega)^2-\mathbf q^2\right]\left[(M/2\pm\omega)-\mathbf q^2\right]}
\\\notag
-&\frac{2M^2}{(M/2\mp \omega)^2-\mathbf q^2}-\frac{M^2}{(M/2\pm \omega)^2-\mathbf q^2}
\bigg]
\\\notag
\times&
\left[1+2 f_A(|(q_{1\pm}-k_+)\cdot u|)\right]
f_\ell((p-k_+)\cdot u) f_\phi(k_+\cdot u)
\,,
\end{align}
where
\begin{align}
{q_{\rm max,min}}=\left|\frac M2 \pm \sqrt{\omega^2-\lambda^2}\right|
\,.
\end{align}
Here, we parametrise the angular dependences as
\begin{align}
\mathbf q=|\mathbf q|\left(\begin{array}{c}\sin\vartheta\sin\varphi\\\sin\vartheta\cos\varphi\\\cos\vartheta\end{array}\right)
\,,\quad
\mathbf k=|\mathbf k|\left(\begin{array}{c}0\\0\\1\end{array}\right)
\,,\quad
\mathbf u=\frac{|\tilde{\mathbf{p}}|}{M}\left(\begin{array}{c}\sin\varrho\sin\psi\\ \sin\varrho\cos\psi\\ \cos\varrho\end{array}\right)
\,.
\end{align}
In order to evaluate the distribution functions, we use
\begin{subequations}
\label{ver:gauge:exponents}
\begin{align}
\label{ver:gauge:exponents:a}
k_+\cdot u&= \frac 12 (\tilde p_0-|\tilde{\mathbf{p}}|\cos\varrho) = E_-\,,
\\
\label{ver:gauge:exponents:b}
(p-k_+)\cdot u&=\frac 12 (\tilde p_0+|\tilde{\mathbf{p}}|\cos\varrho) = E_+\,,
\\
q_{1\pm} \cdot u&=\tilde p_0\left(\frac 12 \pm \frac\omega M\right)
-|\tilde{\mathbf{p}}|\frac {|\mathbf q|}{M}\left(\cos\vartheta\cos\varrho+\sin\vartheta\sin\varrho\cos(\varphi-\psi)\right)\,,
\\
\cos \vartheta &=\frac{\mathbf q^2+\frac{M^2}{4}-\omega^2+\lambda^2}{M|\mathbf q|}\,.
\end{align}
\end{subequations}

Next, we change variables to
\begin{align}
\label{subst:qw:to:xy}
x&=-\frac{1}{2M}\left[\left(\frac M2 -\omega\right)^2-\mathbf q^2\right]\,,
\qquad
y=\frac{1}{2M}\left[\left(\frac M2 +\omega\right)^2-\mathbf q^2\right]\,.
\end{align}
Notice that $x=0$ corresponds to the pole of the lepton (Higgs) propagator
for $\pm\to+$ ($\pm\to-$) and that vice versa, $y=0$ corresponds to the pole of the lepton (Higgs) propagator
for $\pm\to-$ ($\pm\to+$).
The inverse transformation gives
\begin{subequations}
\begin{align}
|\mathbf q|&=\frac12\sqrt{M^2+4M(x-y)+4(x+y)^2}\,,
\\
\omega&=x+y\,,
\end{align}
\end{subequations}
and the Jacobian is $-M/|\mathbf q|$.
We express the vertex contributions as (we take $\lambda \to 0$ at this point):
\begin{align}
\label{Sigmaver:xy}
{\rm tr}[\slashed p {\rm i} \slashed \Sigma^<(p)]^{\rm vert}_i
=&4 g_w G Y^2\frac{1}{32(2\pi)^4}
\int\limits_0^{2\pi}d(\varphi-\psi)\int\limits_{-1}^{1}d\cos\varrho
\int_0^\infty d y \int_0^\infty dx
\\\notag
&\times\left({\cal J}^{\rm vert}_{i+}(x,y)+{\cal J}^{\rm vert}_{i-}(x,y)\right)\,,
\end{align}
where for the present terms
\begin{align}
\label{J:ver1}
{\cal J}^{\rm vert}_{1\pm}(x,y)=&
K^{\rm vert}_{1\pm}(x,y)
\left[1+2 f_A(|(q_{1\pm}-k_+)\cdot u|)\right]
f_\ell(E_+) f_\phi(E_-)
\,,
\end{align}
with
\begin{align}
\label{K:virt1}
K^{\rm vert}_{1\pm}(x,y)=&-{\rm PV}\left[
\vartheta(\mp)
\frac{M^2+2Mx-My}{xy}
+\vartheta(\pm)
\frac{M^2-2My+Mx}{xy}
\right]\,.
\end{align}

In this form, we can directly compare with the terms from
real emissions.
When either or both, $x$, $y$ are close to zero, we can approximate
\begin{align}
|\mathbf q|&\approx\frac12|M+2x-2y|\,,
\\
\label{ver1:arguments}
(q_{1\pm}-k_+)\cdot u&\approx\pm\tilde p_0\frac{x+y}{M}
-|\tilde{\mathbf{p}}|\cos\varrho\frac{x-y}{M}
-2\frac{\sqrt{xy}}{M}|\tilde{\mathbf{p}}|\sin\varrho\cos(\varphi-\psi)\,.
\end{align}

Expanding the arguments of the distribution functions close
to the surfaces $x=0 \,\dot\lor\, y=0$, we obtain
\begin{align}
{\cal J}^{\rm vert}_{1\pm}(x,y)\approx
K^{\rm vert}_{1\pm}(x,y)
\left[1+2 f_A\left(\frac{2x}{M}E_\mp + \frac{2y}{M}E_\pm\right)\right]f_\ell(E_+)f_\phi(E_-)\,.
\end{align}

\subsection{Vertex Correction with on-Shell Higgs Boson}
\label{sec:vertex:virt:2}

 We now consider the third term of Eq.~(\ref{Sigmaless:start}), which using
 the reparametrisation~(\ref{reparametrisation:vertex}) becomes
\begin{align}
{\rm tr}[\slashed p  {\rm i}\slashed \Sigma_N^<&(p)]^{\rm vert}_2
=4 g_w G Y^2\int\frac{d^4 k}{(2\pi)^4}\frac{d^4 q}{(2\pi)^4}
[-k-q]^\nu{\rm tr}\Big[\slashed p
\\\notag
\times&{\rm i}S_\ell^H(p-q)\gamma^\mu
{\rm i}S_\ell^<(p-k)
{\rm i}\Delta_\phi^<(k)
{\rm i}\Delta_\phi^F(q)
{\rm i}\Delta^H_{\mu\nu}(q-k)
\Big]\,.
\end{align}
Making use of the on-shell delta-functions, we find
\begin{align}
\label{Sigma:virt2:omega}
{\rm tr}[\slashed p {\rm i} \slashed \Sigma_N^<(p)]^{\rm vert}_2
=&
4 g_w G Y^2\frac{1}{32(2\pi)^4}
\sum\limits_\pm\int_0^{2\pi}d(\varphi-\psi)\int\limits_{-1}^{1} d\cos\varrho
\int\limits_\lambda^\infty d\omega \int\limits_{q_{\rm min}}^{q_{\rm max}}dq\frac{\omega}{M}
\\\notag
\times&
{\rm PV}\bigg[
\frac{4M^3}{(M\pm \sigma 2 q)(M/2\pm \sigma q+\omega)(M/2\pm \sigma q-\omega)}
\\\notag
+&\frac{4M}{M\pm \sigma 2 q}-\frac{2M^2}{(M/2\pm \sigma q+\omega)(M/2\pm \sigma q-\omega)}
\bigg]
\\\notag
\times&
\left[1+2 f_\phi(|(q_{2\pm\sigma}-k_+)\cdot u|)\right]
f_\ell((p-k_+)\cdot u) f_\phi(k_+\cdot u)
\,,
\end{align}
where we define
\begin{align}
\sigma={\rm sign}\left(\frac{M}{2}-y\right)\,.
\end{align}
Note that in the expression~(\ref{Sigma:virt2:omega}), $q$ is a real number, not a
four-vector. In terms of the original integration variables, the new variables are given
here by
\begin{align}
\omega=\sqrt{q^2+\lambda^2}\,,\quad q=\sqrt{(\mathbf k -\mathbf q)^2}\,.
\end{align}

We again change variables according to Eqs.~(\ref{subst:qw:to:xy})  and with
$|\mathbf q|\to q$ and
decompose the present contribution as in Eqs.~(\ref{Sigmaver:xy}). Notice that the
IR divergences only occur for the contributions where $\pm\to -$,
such that apart from the Bose divergence in the Higgs boson distribution function
${\rm tr}[\slashed p  {\rm i}\slashed \Sigma_N^<(p)]^{\rm vert}_2$ is finite.
The integrand for Eq.~(\ref{Sigmaver:xy}) with $i=2$ then is
\begin{align}
\label{J:ver2}
{\cal J}^{\rm vert}_{2\pm}(x,y)=&
K^{\rm vert}_{2\pm}(x,y)
\left[1+2 f_\phi(|(q_{2\pm\sigma}-k_+)\cdot u|)\right]
f_\ell(E_+)f_\phi(E_-)
\,,
\end{align}
where $k_+\cdot u$ and $(p-k_+)\cdot u$ are again given by Eq.~(\ref{ver:gauge:exponents}) and
\begin{align}
\label{K:virt2}
K^{\rm vert}_{2\pm}(x,y)=&\frac{\omega}{q}
{\rm PV}\Bigg[
\frac{4 M^3}{(M\pm\sigma 2 q)(M/2\pm\sigma q+\omega)(M/2\pm\sigma q-\omega)}+\frac{4 M}{M\pm\sigma 2q}
\\\notag
-&\frac{2M^2}{(M/2\pm\sigma q+\omega)(M/2\pm \sigma q-\omega)}
\Bigg]\,.
\end{align}
We do not explicitly substitute $x$ and $y$ here for $q$ and $\omega$ as this
would give rise to a lengthy expression in this case.

In order to compare with the other contributions, we notice that
when either or both, $x$ and $y$ are small, we may approximate
\begin{subequations}
\label{onshellHiggs:coll:appr}
\begin{align}
\left(\frac M2-q+\omega\right)\left(\frac M2 -q -\omega\right)
\approx& -4xy\,,
\\
M-2q\approx&2(y-x)\,.
\end{align}
\end{subequations}

The relations for constructing the arguments of the distribution functions are
\begin{subequations}
\label{onshellHiggs:distr:arguments}
\begin{align}
q_{2\pm} \cdot u&=\tilde p_0\left(\frac 12 \pm \frac qM\right)
-|\tilde{\mathbf{p}}|\frac{\sqrt{\omega^2-\lambda^2}}{M}\left(\cos\vartheta\cos\varrho+\sin\vartheta\sin\varrho\cos(\varphi-\psi)\right)\,,
\\
\cos \vartheta &=\frac{\omega^2+\frac{M^2}{4}-q^2-\lambda^2}{M\sqrt{\omega^2-\lambda^2}}\,.
\end{align}
\end{subequations}
When either $x$, $y$ or both are close to zero, we may approximate (again, we take $\lambda \to 0$ here)
\begin{align}
\label{onshellHiggs:coll:appr:distrarg}
(q_{2\pm}-k_+)\cdot u\approx&\left(\frac 12+\frac{x-y}{M}\right)\left(\pm\tilde p_0
+|\tilde{\mathbf{p}}|\cos\varrho\right)
-2\frac{\sqrt{xy}}{M}|\tilde{\mathbf{p}}|\sin\varrho\cos(\varphi-\psi)
\,.
\end{align}

Using these approximations, we find for $x\ll M \,\dot\lor\, y\ll M$
\begin{subequations}
\label{J:ver:appr}
\begin{align}
K^{\rm vert,appr}_{2+}(x,y)=&
{\rm PV}\frac{4M(x+y)}{(M+2x-2y)^2}
\,,\\
K^{\rm vert,appr}_{2-}(x,y)=&{\rm PV}\left[\frac{M(M+x-y)}{xy}-\frac{4M}{M+2x-2y}\right]{\rm sign}(x-y)\,,
\\[3mm]
{\cal J}^{\rm vert}_{2\pm}(x,y)\approx&
K^{\rm vert,appr}_{2\pm}(x,y){\rm sign}(M-2y)
\\\notag
\times&
\left[1+2 f_\phi\left(\left|1+\frac{2x-2y}{M}\right|E_\pm\right)\right]f_\ell(E_+)f_\phi(E_-)
\,.
\end{align}
\end{subequations}

\subsection{Vertex Correction with on-Shell Lepton}
\label{sec:vertex:virt:3}

 The third term of Eq.~(\ref{Sigmaless:start}) in
 the reparametrisation~(\ref{reparametrisation:vertex}) is
\begin{align}
{\rm tr}[\slashed p {\rm i} \slashed \Sigma_N^<&(p)]^{\rm vert}_3
=4 g_w G Y^2\int\frac{d^4 k}{(2\pi)^4}\frac{d^4 q}{(2\pi)^4}
[-k-q]^\nu{\rm tr}\Big[\slashed p
\\\notag
\times&{\rm i}S_\ell^F(p-q)\gamma^\mu
{\rm i}S_\ell^<(p-k)
{\rm i}\Delta_\phi^<(k)
{\rm i}\Delta_\phi^H(q)
{\rm i}\Delta^H_{\mu\nu}(q-k)
\Big]
\,.
\end{align}
The kinematic situation is essentially the same as for the on-shell
Higgs boson, which is why the relations~(\ref{onshellHiggs:coll:appr},\ref{onshellHiggs:distr:arguments},\ref{onshellHiggs:coll:appr:distrarg})
apply to the present contribution as well. The integrand can be expressed as
\begin{align}
\label{J:ver3}
{\cal J}^{\rm vert}_{3\pm}(x,y)=&
K^{\rm vert}_3(x,y)
\left[1-2 f_\ell(|(q_{2\pm\sigma} -k_+)\cdot u|)\right]
f_\ell(E_-)f_\phi(E_+)
\,,
\end{align}
where
\begin{align}
\label{K:virt3}
K^{\rm vert}_{3\pm}(x,y)=&\frac{\omega}{q}
{\rm PV}\Bigg[
\frac{4 M^3}{(M\pm\sigma 2 q)(M/2\pm\sigma q+\omega)(M/2\pm\sigma q-\omega)}+\frac{4 M}{M\pm\sigma 2q}
\\\notag
-&\frac{4M^2}{(M/2\pm\sigma q+\omega)(M/2\pm \sigma q-\omega)}
\Bigg]\,.
\end{align}
The arguments for the distribution functions can be constructed using
Eqs.~(\ref{onshellHiggs:distr:arguments}), with the important exception
that we replace $\cos\varrho\to-\cos\varrho$ in order to interchange $E_+ \leftrightarrow E_-$. 
This change of variable leaves the value of the integral unaffected, but it is helpful in order to
to combine the integrands to a manifestly IR-convergent expression
for the right-handed neutrino production rate.

The portion of the
integrand that leads to IR divergences is approximated
for $x\ll M \,\dot\lor\, y\ll M$ as
\begin{align}
K^{\rm vert,appr}_{3+}(x,y)=&0
\,,\\
K^{\rm vert,appr}_{3-}(x,y)=&{\rm PV}\left[\frac{M(M+2x-2y)}{xy}-\frac{4M}{M+2x-2y}\right]{\rm sign}(x-y)\,,
\\[3mm]
{\cal J}^{\rm vert}_{3\pm}(x,y)\approx&
K^{\rm vert,appr}_{3\pm}(x,y){\rm sign}(M-2y)
\\\notag
\times&
\left[1-2 f_\ell\left(\left|1+\frac{2x-2y}{M}\right|E_\mp\right)\right]f_\ell(E_+)f_\phi(E_-)
\,.
\end{align}

\subsection{Infrared Finite Integral}
\label{sec:IR:finite}

The remaining task is to combine the various pieces to an integral, that
can be evaluated with a finite result (in the principal value sense)
for each subset of the complete integration region. This is achieved
by the combination
\begin{subequations}
\begin{align}
\label{integral:total}
{\rm tr}[\slashed p {\rm i} \slashed \Sigma_N^<(p)]^{\rm VERT}
=&4 g_w G Y^2 \frac{1}{32(2\pi)^4}\int\limits_0^{2\pi} d(\varphi-\psi)\int\limits_{-1}^{1}d\cos\varrho
\int\limits_0^\infty dx \int\limits_0^\infty dy
{\cal J}^{\rm total}(x,y)\,,
\\
\label{integrand:IR:finite}
{\cal J}^{\rm total}(x,y)
=&\frac18\Bigg(
{\cal J}^{\rm sca}(x,y)+{\cal J}^{\rm sca}(-x,y)+{\cal J}^{\rm sca}(x,-y)+{\cal J}_{1-}^{\rm vert}(x,y)+{\cal J}_{2-}^{\rm vert}(x,y)
\notag\\\notag
+&
{\cal J}^{\rm sca}(x,y)+{\cal J}^{\rm sca}(x,-y)+{\cal J}^{\rm sca}(-x,y)+{\cal J}_{1+}^{\rm vert}(y,x)+{\cal J}_{3-}^{\rm vert}(y,x)
\\\notag
+&
{\cal J}^{\rm sca}(-y,-x)+{\cal J}_{1+}^{\rm vert}(x,y)+{\cal J}_{2-}^{\rm vert}(x,y)
\\\notag+&
{\cal J}^{\rm sca}(-y,-x)+{\cal J}_{1-}^{\rm vert}(y,x)+{\cal J}_{3-}^{\rm vert}(y,x)
\\+&
+2{\cal J}_{2+}^{\rm vert}(x,y)+2{\cal J}_{3+}^{\rm vert}(y,x)
+x\leftrightarrow y
\Bigg)+\varphi\to\varphi+\pi\,.
\end{align}
\end{subequations}
(Recall that the support of ${\cal J}^{\rm sca}$ is given by Eq.~(\ref{real:support}) 
and shown in Figure~\ref{fig:kinregions}.)
This expression may be integrated without introducing an infrared regulator.

In order to show that this is integrable on the fringes where
$x\to 0 \lor y\to 0$, we make use of the detailed balance relations
applicable to collinear splittings
\begin{subequations}
\label{detailed:balance}
\begin{align}
f_\phi(E)+f_\phi(E)f_\phi(\alpha E)+f_\phi(E)f_A((1-\alpha)E)=&f_\phi(\alpha E)f_A((1-\alpha)E)\,,
\\
f_\ell(E)-f_\ell(E)f_\ell(\alpha E)+f_\ell(E)f_A((1-\alpha)E)=&f_\ell(\alpha E)f_A((1-\alpha)E)\,.
\end{align}
\end{subequations}
It then follows that
for $x\to0$ (divergences in the Higgs boson [lepton] propagator for
Eq.~(\ref{cancel:x:A}) [Eq.~(\ref{cancel:x:B})])
\begin{subequations}
\label{cancel:x}
\begin{align}
\label{cancel:x:A}
xy\left({\cal J}^{\rm sca}(x,y)+{\cal J}^{\rm sca}(-x,y)+{\cal J}_{1-}^{\rm vert}(x,y)+{\cal J}_{2-}^{\rm vert}(x,y)\right) \to\, &0\,,
\\
\label{cancel:x:B}
xy\left({\cal J}^{\rm sca}(-y,-x)+{\cal J}_{1-}^{\rm vert}(y,x)+{\cal J}_{3-}^{\rm vert}(y,x)\right) \to\, &0
\end{align}
\end{subequations}
and that for $y\to0$ (divergences in the lepton [Higgs boson] propagator
for Eq.~(\ref{cancel:y:A}) [Eq.~(\ref{cancel:y:B})])
\begin{subequations}
\label{cancel:y}
\begin{align}
\label{cancel:y:A}
xy\left({\cal J}^{\rm sca}(x,y)+{\cal J}^{\rm sca}(x,-y)+{\cal J}_{1+}^{\rm vert}(y,x)+{\cal J}_{3-}^{\rm vert}(y,x)\right) \to\, &0\,,
\\
\label{cancel:y:B}
xy\left({\cal J}^{\rm sca}(-y,-x)+{\cal J}_{1+}^{\rm vert}(x,y)+{\cal J}_{2-}^{\rm vert}(x,y)\right) \to\,&0\,.
\end{align}
\end{subequations}
The sum ${\cal J}^{\rm sca}(x,y)+{\cal J}^{\rm sca}(-x,y)$ in Eq.~(\ref{cancel:x:A}) makes it sure that the cancellation with vertex contributions takes place on the whole collinear edge $x=0$ (for both $y<M/2$ and $y\geq M/2$), and similarly in Eq.~(\ref{cancel:y:A}) for $y=0$ in accordance with the kinematic regions, {\it cf.} Figure~\ref{fig:kinregions}. In order to obtain the manifestly IR-finite result for the vertex-type part
of the neutrino production rate~(\ref{integral:total}), we need to add these
contributions to those terms, that are integrable from the outset and finally
symmetrise $x\leftrightarrow y$, in order to make sure that the sum of all
divergent terms cancels on both edges ($x=0$ and $y=0$). 
Furthermore, we have symmetrised over the polar angles.

The criteria~(\ref{cancel:x},\ref{cancel:y}) are in general not sufficient in order
to show the integrability at the point $x=y=0$, because the detailed balance relations
only hold on the collinear fringes where $x=0\,\dot\lor\,y=0$. We must therefore
show in addition that $xy{\cal J}^{\rm total}(x,y)$ vanishes on each trajectory
for $(x,y) \to 0$. The problematic terms are the Bose divergences
in ${\cal J}^{\rm sca}$ and ${\cal J}_{1\pm}^{\rm vert}$. When using
Eqs.~(\ref{J:sca},\ref{J:ver1}) together with the
approximations for the arguments of the distribution functions~(\ref{scat:arguments},\ref{ver1:arguments}), we notice the cancellation
of the divergences in $xy({\cal J}^{\rm sca}(x,y)+{\cal J}_{1\pm}^{\rm vert}(x,y))$. There
is however a finite remainder that moreover depends on the direction of approach of
$(x,y)$ toward $(0,0)$. This finite contribution follows from the terms that
are linear in $x$, $y$ and $\sqrt{xy}$ in the arguments of the distribution 
functions~(\ref{scat:arguments},\ref{ver1:arguments}). For this finite term to cancel as well, we must add
${\cal I}^{\rm sca}(-x,-y)$, symmetrise over $x$ and $y$ (as it is already necessary
for the cancellation of the collinear divergences) and moreover, add the contributions
from positive and negative values of $\cos(\varphi-\psi)$. In summary,
it follows that 
\begin{align}
\lim\limits_{x,y\to 0}xy{\cal J}^{\rm total}(x,y)=0\,,
\end{align}
 such that ${\cal J}^{\rm total}(x,y)$ grows more
slowly than $1/(x^2+y^2)$, and therefore it is
 integrable at $x=y=0$. In fact, we have verified by numerical evaluation that ${\cal J}^{\rm total}(x,y)\sim 1/\sqrt{x^2+y^2}$ for $(x,y)\to 0$.

Similarly, care must be taken about the points $(x,y)=(0,M/2)$ and $(x,y)=(M/2,0)$,
where the Bose divergence of the Higgs scalar coincides with the collinear fringes.
Suppose we are close to the fringe where $x\to 0$. Inspecting the
approximation~(\ref{J:ver:appr}) for ${\cal J}^{\rm vert}_{2\pm}$,
we notice that the $y$ integration can be performed
at $y=M/2$ in the principal value sense. The same is true for
the collinearly divergent terms $\propto 1/x$ in
${\cal J}^{\rm sca}({\rm sign}(M/2-y)x,y)$, Eq.~(\ref{J:sca}).
Therefore, we can perform the $d \Delta y$ integral over
${\cal J}^{\rm total}(x,M/2-\Delta y)+{\cal J}^{\rm total}(x,M/2+\Delta y)$
as well ($\Delta y>0$).
Due to the detailed balance relations~(\ref{cancel:x},\ref{cancel:y}), $x$ times this integral
vanishes for $x\to 0$, which is why the remaining $x$ integration yields a finite
result as well. We have also checked numerically that for $(x,y)\to (0,M/2)$
\begin{align}
 x(y-M/2){\cal J}^{\rm total}(x,y) \to\,0\,,
\end{align}
and moreover that ${\cal J}^{\rm total}(x,y)\sim 1/\sqrt{x^2+(y-M/2)^2}$ irrespective of the limiting direction, which further confirms the integrability at $(0,M/2)$ and $(M/2,0)$ by  $x \leftrightarrow y$ symmetry. Furthermore, we have verified that the Bose divergences in ${\cal J}^{\rm vert}_{2\pm}$ on the line $y = M/2 + x$ away from fringe $x=0$ are actually cancelled in that $(y-M/2-x){\cal J}^{\rm total}(x,y) \to 0$ and ${\cal J}^{\rm total}(x,y) \to {\rm finite}$ for $y \to M/2 + x$ , and similarly for $x \leftrightarrow y$.

\subsection{Ultraviolet Regularisation}
\label{sec:vert:UV}

The vertex-type correction~(\ref{integral:total}) to the neutrino production contains
UV divergences that need to be isolated and subtracted in order to obtain a remainder that
is suitable for numerical integration. The UV divergences originate from the vertex
contributions. In particular, the integrands~(\ref{J:ver1},\ref{J:ver2},\ref{J:ver3})
are not exponentially suppressed for large values of $x$ or $y$ because of
the constant term in the square brackets. The contributions arising from
this constant term are easily interpreted as the vacuum vertex-corrections to the
$2\to 1$ processes $\ell,\phi\to N$ or $\bar\ell\phi^\dagger\to N$,
weighted by the statistical distributions for $\ell$ and $\phi$.
These UV-divergent contributions are analytically calculable, and are given by
the the following sum of terms:
\begin{subequations}
\begin{align}
{\cal J}^{\rm vert,vac}(x,y)=&
\sum\limits_{\pm}\left(K^{\rm vert}_{1\pm}(x,y)+K^{\rm vert}_{2\pm}(x,y)+K^{\rm vert}_{3\pm}(x,y)\right)
f_\ell(E_+)f_\phi(E_-)\,.
\end{align}
\end{subequations}
Now, we need to subtract this expression from the integrand that we aim to evaluate
numerically, while adding terms that balance the collinear divergences that this
subtraction may induce.
The vacuum-vertex corrections should only have IR divergences that cancel
with those for real emissions in $1\to3$ processes. (Scatterings of the
$2\leftrightarrow2$ type do not occur in the vacuum,
in absence of incoming gauge bosons, leptons
and Higgs bosons.)
This implies that outside of Region~I, ${\cal J}^{\rm vert}_{\overline{\rm UV}}(x,y)$
should not yield collinearly divergent contributions, which can be verified as
\begin{align}
xy({\cal J}^{\rm vert,vac}(x,y)+{\cal J}^{\rm vert,vac}(y,x))=0\quad\textnormal{for}\quad(x=0 \land y > M/2) \lor (y=0 \land x> M/2)\,.
\end{align}
This is no longer true for Region~I. In order to cancel collinear divergences there,
we define
\begin{align}
{\cal J}^{\rm sca,vac}(x,y)=K^{\rm sca}(x,y)
\vartheta(x+y-M/2)f_\ell(E_+)f_\phi(E_-)\,,
\end{align}
which we subtract in addition. This is just $1\to 3$ rate weighted with the same statistical
factors for leptons and Higgs bosons
as ${\cal J}^{\rm vert,vac}(x,y)$, and it has the desired property
that
\begin{align}
xy({\cal J}^{\rm sca,vac}(x,y)+{\cal J}^{\rm vert,vac}(x,y)+x\leftrightarrow y)
=0\qquad\textnormal{for}\;x=0  \lor y=0\,.
\end{align}
Now the integrand
\begin{align}
{\cal J}^{\rm total}_{\overline{\rm UV}}(x,y)=
{\cal J}^{\rm total}(x,y)-\frac 12\left({\cal J}^{\rm sca,vac}(x,y)+{\cal J}^{\rm vert,vac}(x,y)+x\leftrightarrow y\right)
\end{align}
is IR and UV finite and suitable for numerical evaluation.

Finally, the subtracted terms must be calculated analytically and then
added again. For this purpose, we note that
the virtual $T=0$ corrections to the decay rate, evaluated with a cutoff
$\omega<\Lambda$ [from the parametrisations in Eqs.~(\ref{Sigma:virt1:omega},\ref{Sigma:virt2:omega})] are
\begin{align}
{\rm tr}[\slashed p\Sigma^{\cal A}_N(p)]^{\rm vert}_{T=0}
=&\frac{4 g_w G Y^2}{16(2\pi)^3}
M^2\left[
-2\log^2\frac{\lambda}{M}+\frac{\pi^2}{6}-3\log\frac{\lambda}{M}+\log\frac{\Lambda}{M}-\frac32+\log2
\right]
\\\notag
+&Y\delta Y\frac{M^2}{4\pi}
\,.
\end{align}
Here, $\delta Y$ is a counter term that must remove the logarithmic 
divergence in $\Lambda$ and the further details of which are
determined by a particular renormalisation scheme.

Alternatively, we can use dimensional regularisation in order to calculate the
reduced vertex function
\begin{align}
V(p)=&2 g_w G Y^2
\mu^\epsilon
\int\frac{d^{4-\epsilon} q}{(2\pi)^{4-\epsilon}}
\frac{\frac12{\rm tr}[\slashed p(\slashed p-\slashed q) \gamma_\mu (\slashed p-\slashed k)](k+q)^\mu}{[(p-q)^2+{\rm i}\varepsilon][(q-k)^2-\lambda^2+{\rm i}\varepsilon][q^2+{\rm i}\epsilon]}
+Y\delta Y 2 M^2
\\\notag
=&2 g_w G Y^2 \frac{M^2}{8\pi^2}\left[
-2\log^2\frac{\lambda}{M}+\frac{\pi^2}{6}-3\log\frac{\lambda}{\mu}+2\log\frac{M}{\mu}
-\Delta_\epsilon-\frac 12
\right]+Y\delta Y2 M^2\,.
\end{align}
In this form, $\delta Y$ must cancel the $1/\epsilon$ divergence and may
otherwise again be renormalisation-scheme dependent.
Integrating over the $T=0$ phase space, we obtain the zero-temperature
spectral-function for the right-handed neutrino
\begin{align}
{\rm tr}[\slashed p\Sigma^{\cal A}_N(p)]^{\rm vert}_{T=0}
=&\int\frac{d^4 k}{(2\pi)^4}\frac{d^4 q}{(2\pi)^4}
(2\pi)^4\delta^4(p-k-q)
2\pi\delta(k^2)2\pi\delta(q^2)V(p)
=\frac{V(p)}{8\pi}
\\\notag
=&
\frac{4 g_w G Y^2}{2^7\pi^3}
M^2\left[
-2\log^2\frac{\lambda}{M}+\frac{\pi^2}{6}-3\log\frac{\lambda}{\mu}+2\log\frac{M}{\mu}
-\Delta_\epsilon-\frac 12
\right]
\\\notag
+&Y\delta Y\frac{M^2}{4\pi}\,.
\end{align}
Note the factors from the two different diagrammatic contributions in the CTP approach
(cut on the left or on the right).
Comparing with the prefactors in Eq.~(\ref{Sigmaver:xy}) and using the
definitions~(\ref{CTP:combinations}), we can conclude that
\begin{subequations}
\begin{align}
\label{J:vert:vac}
\bar{\cal J}^{\rm vert,vac}=&
\int\limits_0^\infty dx \int\limits_0^\infty dy
{\cal J}^{\rm vert,vac}(x,y)\,,
\\
\bar{\cal J}^{\rm vert,vac}(x,y)=&
-2\,M^2\left[
-2\log^2\frac{\lambda}{M}+\frac{\pi^2}{6}-3\log\frac{\lambda}{\mu}+2\log\frac{M}{\mu}
-\Delta_\epsilon-\frac 12
\right]
\\\notag
&\times
f_\ell(E_+)f_\phi(E_-)\,.
\end{align}
\end{subequations}
The IR divergences are again isolated in terms involving $\log\lambda$,
which cancel with corresponding terms in the vacuum
vertex contribution~(\ref{J:vert:vac}).

Next, we need to obtain a matching expression for
${\cal J}^{\rm sca,vac}(x,y)$, which we have introduced in order
to cancel the IR-divergent $\lambda$-dependence. For this purpose,
we evaluate the integral
\begin{align}
\int\limits_{\frac{\lambda^2}{2M}}^{\frac M2}dx
\int\limits_{\frac{\lambda^2}{4x}}^{\frac M2-x}dy
\left[
-2\frac{M^2}{xy}+4\frac My+2\frac Mx +4
\right]
=2 M^2\left[-2\log^2\frac{\lambda}{M}+\frac{\pi^2}{6}+3\log\frac{M}{\lambda}-\frac{11}{2} M^2\right]\,,
\end{align}
where the boundaries of integration derive from the condition
$-1\leq\cos\vartheta\leq 1$ with Eq.~(\ref{rel:costheta}).
Hence,
\begin{align}
\label{J:sca:vac}
\bar{\cal J}^{\rm sca,vac}=&
\int\limits_0^\infty dx \int\limits_0^\infty dy
{\cal J}^{\rm sca,vac}(x,y)
\\\notag
=&
2\,M^2\left[
-2\log^2\frac{\lambda}{M}+\frac{\pi^2}{6}+3\log\frac{M}{\lambda}-\frac{11}{2} M^2
\right]
f_\ell(E_+)f_\phi(E_-)\,.
\end{align}

The final expression for the reduced self-energy is
\begin{align}
{\rm tr}[\slashed p {\rm i} \slashed \Sigma_N^<(p)]^{\rm VERT}
=&4 g_w G Y^2 \frac{1}{32(2\pi)^4}\int\limits_0^{2\pi} d(\varphi-\psi)\int\limits_{-1}^{1}d\cos\varrho
\int\limits_0^\infty dx \int\limits_0^\infty dy{\cal J}^{\rm total}_{\overline{\rm UV}}(x,y)
\\\notag
+&4 g_w G Y^2 \frac{1}{32(2\pi)^3}
\int\limits_{-1}^{1}d\cos\varrho
\left(
\bar{\cal J}^{\rm vert,vac}+\bar{\cal J}^{\rm sca,vac}
\right)
\\\notag
+&
\frac{M^2}{4\pi}Y\delta Y\int\limits_{-1}^{1}d\cos\varrho\,
f_\ell(E_+)f_\phi(E_-)
\,.
\end{align}
Each of the explicit integrals in this expression is convergent, while
the remaining UV and IR divergences are isolated within
simple factors that cancel in the sum,
as discussed above in this present Section.

\section{Discussion and Conclusions}
\label{section:conclusions}

The results of this work show that the relaxation rate of
right-handed neutrinos $N$ toward thermal equilibrium is finite to
NLO in perturbation theory, {\it i.e.} when including
the leading corrections from Standard Model gauge radiation. As we perform a
fully relativistic calculation, we generalise earlier results
that are based on non-relativistic approximations~\cite{Salvio:2011sf,Laine:2011pq}
that are valid when
$M\gg T$. In particular, the relative motion of the neutrino $N$ with respect to
the plasma is kept general here and the full quantum statistical Bose-Einstein and
Fermi-Dirac effects are accounted for.

We have developed two somewhat different methods in order to handle the wave-function-
and the vertex-type corrections. The treatment of the wave-function corrections
is greatly facilitated, because well known and relatively simple analytical
expressions for the thermal self-energies are available. We can isolate the
IR divergences in terms of
logarithmic dependences on the regulating gauge-boson mass $\lambda$ for
both, the scattering contributions that rely on the spectral self-energy
and the wave-function contributions originating from the hermitian
self-energy. Eventually, we find that the IR divergences cancel
[Eq.~(\ref{B:virt:vac}) with Eq.~(\ref{B:real:vac+}), Eq~(\ref{F:col:wv:vac}) with Eq.~(\ref{F:col:sca:vac}) and Eq~(\ref{F:HTL:col:wv}) with Eq.~(\ref{F:HTL:col:sca})]
because the
hermitian and the spectral self-energies are the real and imaginary parts
of the same analytic function evaluated at the two-particle branch cut.

For the two-particle irreducible vertex-type corrections, the situation appears more
complicated. Rather than isolating the IR divergences in analytic expressions for
various contributions, we therefore manipulate these in such a manner that we
obtain the integrand~(\ref{integrand:IR:finite})
that is manifestly free from non-integrable IR and Bose
divergences.

Since the methods that we use in order to demonstrate the cancellation are
very explicit and as we also discuss the subtraction of UV divergences,
they readily suggest a method for analytic or numerical evaluation of the
relaxation rates of right-handed neutrinos $N$. In a future work,
it would be of interest to verify the
results of Refs.~\cite{Salvio:2011sf,Laine:2011pq} by taking
the non-relativistic limit of the present results and moreover, to
perform a numerical evaluation
of the production rate of $N$ without non-relativistic approximations.

While the present work provides a method for evaluating the relaxation rate
of right-handed neutrinos $N$, that derives from
${\rm tr}[\slashed p\slashed\Sigma^{\cal A}_N]$, {\it cf.} Eq.~(\ref{kineq:N}),
the rates of non-equilibrium $CP$-violation from decays and inverse
decays of $N$ typically
rely on different components of the
self energy~\cite{Beneke:2010wd,Garbrecht:2011aw,Garbrecht:2012qv,Drewes:2012ma,Garbrecht:2012pq}. Another future task would therefore be the evaluation of $CP$-violating rates,
in generalisation of the methods presented here.

While we have focused in the present work on the production of right-handed neutrinos,
the basic topology of the self-energy diagrams that describe the decay and scattering
rates and their NLO corrections is obviously shared with diagrams
for other high-energy reactions in finite-temperature backgrounds.
For that
reason, it may be useful to formulate our method in terms of master integrals in such a way that the results could be more directly applied to other situations with different kinds of particles in the loops (different Lorentz tensor structures).
One may investigate for example the processes that underlay the transport coefficients used in calculations for Electroweak Baryogenesis or
corrections to the production rate of Dark Matter particles.
It will therefore be interesting to explore the possibilities of applying the present
methods to additional topics in Early Universe Cosmology.

\section*{Acknowledgements}

The authors acknowledge support by the Gottfried Wilhelm Leibniz programme of the
DFG, by the DFG cluster of excellence `Origin and Structure
of the Universe' and by the Alexander von Humboldt
 Foundation.

\renewcommand{\theequation}{A\arabic{equation}}
\setcounter{equation}{0}
\section*{Appendix}

We follow the conventions of Ref.~\cite{Prokopec:2003pj}.
The relations between the CTP two-point functions,
($G$ stands for a propagator $\Delta$ or $S$
or for a self-energy $\Pi$ or $\slashed\Sigma$)
are given by:
\begin{subequations}
\label{CTP:combinations}
\begin{align}
\label{CTP:advanced}
G^A=G^T-G^>=G^<-G^{\bar T}\quad&\textnormal{(advanced)}\,,
\\
\label{CTP:retarded}
G^R=G^T-G^<=G^>-G^{\bar T}\quad&\textnormal{(retarded)}\,,
\\
\label{CTP:hermitian}
G^H=\frac12(G^R+G^A)\quad&\textnormal{(Hermitian)}\,,
\\
\label{CTP:spectral}
G^{\cal A}=\frac1{2\rm i}(G^A-G^R)=\frac{\rm i}2(G^>-G^<)\quad&\textnormal{(anti-Hermitian, spectral)}\,,
\\
\label{CTP:statistic}
G^{F}=\frac12 (G^>+G^<)=\frac12(G^T+G^{\bar T})\quad&\textnormal{(statistic)}\,.
\end{align}
\end{subequations}
In terms of the functions bearing CTP indices, these can be expressed as
\begin{subequations}
\label{CTP:prop:basic}
\begin{align}
G^T=&G^{++}\,,\quad G^{\bar T}=G^{--}\,\quad\textnormal{(time ordered, anti time-ordered)}\,,
\\
\quad G^<=&G^{+-}\,,\quad G^>=G^{-+}\,\quad\textnormal{(Wightman)}\,.
\end{align}
\end{subequations}

The Wigner transform is defined by
\begin{align}
\label{Wigner:transform}
G(p,x)=\int d^4 r {\rm e}^{{\rm i}pr} G(x+r/2,x-r/2)\,.
\end{align}
Under spatially homogeneous conditions, there is no dependence on
the average spatial coordinate $\mathbf x$ and
furthermore, we suppress the explicit average time coordinate $t=x^0$. Equilibrium
Green functions or self energies observe the Kubo-Martin-Schwinger
(KMS) relation
\begin{align}
\label{KMS}
G^{>}(p)=\pm{\rm e}^{\beta p \cdot u}G^{<}(p)\,,
\end{align}
where the plus sign applies to bosonic, the minus sign to fermionic
two-point functions, and $u$ is the plasma four-velocity.

The tree-level equilibrium
propagators for the lepton doublet that we use here are
\begin{subequations}
\label{prop:ell:expl}
\begin{align}
{\rm i}S_{\ell}^{<}(p)
&=-2\pi\delta(p^2){\rm sign}(p^0)P_{\rm L}p\!\!\!/
f_{\ell}(p\cdot u)\,,\\
{\rm i}S_{\ell}^{>}(p)
&=2\pi\delta(p^2){\rm sign}(p^0)P_{\rm L}p\!\!\!/
(1-f_{\ell}(p\cdot u))\,,\\
{\rm i}S_{\ell}^{T}(p)
&=
P_{\rm L}\frac{{\rm i}p\!\!\!/}{p^2+{\rm i}\varepsilon}
-2\pi\delta(p^2)P_{\rm L}p\!\!\!/f_{\ell}(|p\cdot u|)\,,
\\
{\rm i}S_{\ell}^{\bar T}(p)
&=
-P_{\rm L}\frac{{\rm i}p\!\!\!/}{p^2-{\rm i}\varepsilon}
-2\pi\delta(p^2)P_{\rm L}p\!\!\!/
f_{\ell}(|p^0|)
\,,
\end{align}
\end{subequations}
where $f_\ell$ is the Fermi-Dirac distribution. The generalisation to
non-equilibrium distributions can be found in Ref.~\cite{Beneke:2010wd}.
We suppress explicit ${\rm SU}(2)$ indices, which are contracted with that
of the Higgs doublet $\phi$ in the usual manner, using the
two-dimensional anti-symmetric tensor.

Similarly, we use the scalar equilibrium propagators
\begin{subequations}
\label{prop:phi:expl}
\begin{align}
{\rm i}\Delta_\phi^<(p)&=
2\pi \delta(p^2){\rm sign}(p^0)f_\phi(p\cdot u)
\,,
\\
{\rm i}\Delta_\phi^>(p)&=
2\pi \delta(p^2){\rm sign}(p^0)(1+f_\phi(p\cdot u))
\,,
\\
{\rm i}\Delta_\phi^T(p)&=
\frac{\rm i}{p^2+{\rm i}\varepsilon}+
2\pi \delta(p^2)f_\phi(|p\cdot u|)
\,,
\\
{\rm i}\Delta_\phi^{\bar T}(p)&=
-\frac{\rm i}{p^2-{\rm i}\varepsilon}+
2\pi \delta(p^2)f_\phi(|p\cdot u|)
\,,
\end{align}
\end{subequations}
where $f_\phi$ is a Bose-Einstein distribution. The gauge field propagators are
\begin{subequations}
\label{prop:A:expl}
\begin{align}
{\rm i}\Delta_{\mu\nu}^<(p)&=
2\pi \delta(p^2){\rm sign}(p^0)(-g_{\mu\nu})f_A(p\cdot u)
\,,
\\
{\rm i}\Delta_{\mu\nu}^>(p)&=
2\pi \delta(p^2){\rm sign}(p^0)(-g_{\mu\nu})(1+f_A(p\cdot u))
\,,
\\
{\rm i}\Delta_{\mu\nu}^T(p)&=
\frac{-{\rm i}g_{\mu\nu}}{p^2+{\rm i}\varepsilon}+
2\pi \delta(p^2)(-g_{\mu\nu})f_A(|p\cdot u|)
\,,
\\
{\rm i}\Delta_{\mu\nu}^{\bar T}(p)&=
-\frac{-{\rm i}g_{\mu\nu}}{p^2-{\rm i}\varepsilon}+
2\pi \delta(p^2)(-g_{\mu\nu})f_A(|p\cdot u|)
\,,
\end{align}
\end{subequations}
and $f_A$ is a Bose-Einstein distribution.

As a consequence of the Majorana condition,
the self energy of the right-handed neutrinos inherits the property
\begin{align}
\label{Majorana:selfenergy}
\slashed \Sigma_{N}(x,y)=C \slashed \Sigma_{N}^t(y,x) C^\dagger\,,
\end{align}
where the transposition $t$ acts here on both the CTP and the Dirac indices,
{\it cf.} Ref.~\cite{Garbrecht:2011aw}
for a more detailed discussion that also includes the mixing of several
right-handed neutrinos
and the effect of non-zero chemical potentials for the leptons $\ell$.


\begin{thebibliography}{99}
\bibliographystyle{unsrt}

\bibitem{Arnold:2000dr}
  P.~B.~Arnold, G.~D.~Moore and L.~G.~Yaffe,
  ``Transport coefficients in high temperature gauge theories. 1. Leading log results,''
  JHEP {\bf 0011} (2000) 001
  [hep-ph/0010177].

\bibitem{Arnold:2001ms}
  P.~B.~Arnold, G.~D.~Moore and L.~G.~Yaffe,
  ``Photon emission from quark gluon plasma: Complete leading order results,''
  JHEP {\bf 0112} (2001) 009
  [hep-ph/0111107].

\bibitem{Besak:2012qm}
  D.~Besak and D.~Bodeker,
  ``Thermal production of ultrarelativistic right-handed neutrinos: Complete leading-order results,''
  JCAP {\bf 1203} (2012) 029
  [arXiv:1202.1288 [hep-ph]].

\bibitem{Anisimov:2010gy}
  A.~Anisimov, D.~Besak and D.~Bodeker,
  ``Thermal production of relativistic Majorana neutrinos: Strong enhancement by multiple soft scattering,''
  JCAP {\bf 1103} (2011) 042
  [arXiv:1012.3784 [hep-ph]].

\bibitem{Salvio:2011sf}
  A.~Salvio, P.~Lodone and A.~Strumia,
  ``Towards leptogenesis at NLO: the right-handed neutrino interaction rate,''
  JHEP {\bf 1108} (2011) 116
  [arXiv:1106.2814 [hep-ph]].

\bibitem{Laine:2011pq}
  M.~Laine and Y.~Schroder,
  ``Thermal right-handed neutrino production rate in the non-relativistic regime,''
  JHEP {\bf 1202} (2012) 068
  [arXiv:1112.1205 [hep-ph]].


\bibitem{Baier:1988xv}
  R.~Baier, B.~Pire and D.~Schiff,
  ``Dilepton production at finite temperature: Perturbative treatment at order $\alpha_s$,''
  Phys.\ Rev.\ D {\bf 38} (1988) 2814.

\bibitem{Altherr:1989yn}
  T.~Altherr and T.~Becherrawy,
  ``Cancellation Of Infrared And Mass Singularities In The Thermal Dilepton Rate,''
  Nucl.\ Phys.\ B {\bf 330} (1990) 174.

\bibitem{Gabellini:1989yk}
  Y.~Gabellini, T.~Grandou and D.~Poizat,
  ``Electron - Positron Annihilation In Thermal Qcd,''
  Annals Phys.\  {\bf 202} (1990) 436.


\bibitem{Altherr:1988bg}
  T.~Altherr, P.~Aurenche and T.~Becherrawy,
  ``On Infrared And Mass Singularities Of Perturbative Qcd In A Quark - Gluon Plasma,''
  Nucl.\ Phys.\ B {\bf 315} (1989) 436.


\bibitem{Laine:2011xm}
  M.~Laine, A.~Vuorinen and Y.~Zhu,
  ``Next-to-leading order thermal spectral functions in the perturbative domain,''
  JHEP {\bf 1109} (2011) 084
  [arXiv:1108.1259 [hep-ph]].

\bibitem{Zhu:2012be}
  Y.~Zhu and A.~Vuorinen,
  ``The shear channel spectral function in hot Yang-Mills theory,''
  arXiv:1212.3818 [hep-ph].


\bibitem{Schwinger:1960qe}
  J.~S.~Schwinger,
  ``Brownian motion of a quantum oscillator,''
  J.\ Math.\ Phys.\  {\bf 2} (1961) 407.

\bibitem{Keldysh:1964ud}
  L.~V.~Keldysh,
  ``Diagram technique for nonequilibrium processes,''
  Zh.\ Eksp.\ Teor.\ Fiz.\  {\bf 47} (1964) 1515
  [Sov.\ Phys.\ JETP {\bf 20} (1965) 1018].

\bibitem{Calzetta:1986cq}
  E.~Calzetta and B.~L.~Hu,
  ``Nonequilibrium Quantum Fields: Closed Time Path Effective Action, Wigner
  Function and Boltzmann Equation,''
  Phys.\ Rev.\  D {\bf 37} (1988) 2878.

\bibitem{Prokopec:2003pj}
  T.~Prokopec, M.~G.~Schmidt and S.~Weinstock,
  ``Transport equations for chiral fermions to order h bar and electroweak baryogenesis. Part 1,''
  Annals Phys.\  {\bf 314} (2004) 208
  [hep-ph/0312110].

\bibitem{Prokopec:2004ic}
  T.~Prokopec, M.~G.~Schmidt and S.~Weinstock,
  ``Transport equations for chiral fermions to order h-bar and electroweak baryogenesis. Part II,''
  Annals Phys.\  {\bf 314} (2004) 267
  [hep-ph/0406140].

\bibitem{Buchmuller:2000nd} 
  W.~Buchmuller and S.~Fredenhagen,
  ``Quantum mechanics of baryogenesis,''
  Phys.\ Lett.\ B {\bf 483}, 217 (2000)
  [hep-ph/0004145].


\bibitem{De Simone:2007rw}
  A.~De Simone and A.~Riotto,
  ``Quantum Boltzmann Equations and Leptogenesis,''
  JCAP {\bf 0708} (2007) 002
  [hep-ph/0703175].

\bibitem{Garny:2009rv}
  M.~Garny, A.~Hohenegger, A.~Kartavtsev and M.~Lindner,
  ``Systematic approach to leptogenesis in nonequilibrium QFT: vertex
  contribution to the CP-violating parameter,''
  Phys.\ Rev.\  D {\bf 80} (2009) 125027
  [arXiv:0909.1559 [hep-ph]].

\bibitem{Garny:2009qn}
  M.~Garny, A.~Hohenegger, A.~Kartavtsev and M.~Lindner,
  ``Systematic approach to leptogenesis in nonequilibrium QFT: self-energy
  contribution to the CP-violating parameter,''
  Phys.\ Rev.\  D {\bf 81} (2010) 085027
  [arXiv:0911.4122 [hep-ph]].

\bibitem{Anisimov:2010aq}
  A.~Anisimov, W.~Buchm\"uller, M.~Drewes and S.~Mendizabal,
  ``Leptogenesis from Quantum Interference in a Thermal Bath,''
  Phys.\ Rev.\ Lett.\  {\bf 104} (2010) 121102
  [arXiv:1001.3856 [hep-ph]].

\bibitem{Garny:2010nj}
  M.~Garny, A.~Hohenegger, A.~Kartavtsev,
  ``Medium corrections to the CP-violating parameter in leptogenesis,''
  Phys.\ Rev.\  {\bf D81 } (2010)  085028.
  [arXiv:1002.0331 [hep-ph]].

\bibitem{Beneke:2010wd}
  M.~Beneke, B.~Garbrecht, M.~Herranen and P.~Schwaller,
  ``Finite Number Density Corrections to Leptogenesis,''
  Nucl.\ Phys.\  B {\bf 838} (2010) 1
  [arXiv:1002.1326 [hep-ph]].

\bibitem{Beneke:2010dz}
  M.~Beneke, B.~Garbrecht, C.~Fidler, M.~Herranen and P.~Schwaller,
  ``Flavoured Leptogenesis in the CTP Formalism,''
  Nucl.\ Phys.\  B {\bf 843} (2011) 177
  [arXiv:1007.4783 [hep-ph]].

\bibitem{Garny:2010nz}
  M.~Garny, A.~Hohenegger and A.~Kartavtsev,
  ``Quantum corrections to leptogenesis from the gradient expansion,''
  arXiv:1005.5385 [hep-ph].

\bibitem{Garbrecht:2010sz}
  B.~Garbrecht,
  ``Leptogenesis: The Other Cuts,''
  Nucl.\ Phys.\  {\bf B847 } (2011)  350-366.
  [arXiv:1011.3122 [hep-ph]].

\bibitem{Anisimov:2010dk}
  A.~Anisimov, W.~Buchmuller, M.~Drewes and S.~Mendizabal,
  ``Quantum Leptogenesis I,''
  Annals Phys.\  {\bf 326} (2011) 1998
  [arXiv:1012.5821 [hep-ph]].

\bibitem{Garbrecht:2011aw}
  B.~Garbrecht and M.~Herranen,
  ``Effective Theory of Resonant Leptogenesis in the Closed-Time-Path Approach,''
  Nucl.\ Phys.\ B {\bf 861} (2012), 17.
  [arXiv:1112.5954 [hep-ph]].

\bibitem{Garny:2011hg}
  M.~Garny, A.~Kartavtsev and A.~Hohenegger,
  ``Leptogenesis from first principles in the resonant regime,''
  Annals Phys.\  {\bf 328} (2013) 26
  [arXiv:1112.6428 [hep-ph]].


\bibitem{Garbrecht:2012qv}
  B.~Garbrecht,
  ``Leptogenesis from Additional Higgs Doublets,''
  Phys.\ Rev.\ D {\bf 85} (2012) 123509
  [arXiv:1201.5126 [hep-ph]].

\bibitem{Drewes:2012ma}
  M.~Drewes and B.~Garbrecht,
  ``Leptogenesis from a GeV Seesaw without Mass Degeneracy,''
  arXiv:1206.5537 [hep-ph].


\bibitem{Garbrecht:2012pq}
  B.~Garbrecht,
  ``Baryogenesis from Mixing of Lepton Doublets,''
  Nucl.\ Phys.\ B {\bf 868} (2013) 557
  [arXiv:1210.0553 [hep-ph]].

\bibitem{Frossard:2012pc}
  T.~Frossard, M.~Garny, A.~Hohenegger, A.~Kartavtsev and D.~Mitrouskas,
  ``Systematic approach to thermal leptogenesis,''
  arXiv:1211.2140 [hep-ph].


\bibitem{GSG}
  B.~Garbrecht,P.~Schwaller, F.~Glowna, in preparation.

\bibitem{Altherr:1994jc}
  T.~Altherr,
  ``Resummation of perturbation series in nonequilibrium scalar field theory,''
  Phys.\ Lett.\ B {\bf 341} (1995) 325
  [hep-ph/9407249].

\bibitem{Garbrecht:2011xw}
  B.~Garbrecht and M.~Garny,
  ``Finite Width in out-of-Equilibrium Propagators and Kinetic Theory,''
  Annals Phys.\  {\bf 327} (2012) 914
  [arXiv:1108.3688 [hep-ph]].

\bibitem{Millington:2012pf}
  P.~Millington and A.~Pilaftsis,
  ``Perturbative Non-Equilibrium Thermal Field Theory,''
  arXiv:1211.3152 [hep-ph].

\bibitem{Weldon:1982bn}
  H.~A.~Weldon,
  ``Effective Fermion Masses of Order gT in High Temperature Gauge Theories with Exact Chiral Invariance,''
  Phys.\ Rev.\ D {\bf 26} (1982) 2789.

\end{thebibliography}
\end{document}